\documentclass[prd,aps,twocolumn,showpacs,preprintnumbers,amsmath,amssymb]{revtex4}
\usepackage{graphicx}% Include figure files
\usepackage{dcolumn}% Align table columns on decimal point
\usepackage{bm}% bold math
\begin{document}

\title{Heavy hadrons on $N_f=2$ and $2+1$ improved clover-Wilson lattices}
\author{Tommy Burch$^1$
\footnote{tommy.burch@physik.uni-r.de}
}
\vskip1mm
\affiliation{$^1$D-93053 Regensburg, Germany}

\begin{abstract}
We present the masses of singly ($B$, $B_s$, $\Lambda_b$, $\Sigma_b$, etc.), 
doubly ($B_c$, $\eta_b$, $\Upsilon$, $\Xi_{bc}$, $\Xi_{bb}$, etc.), and triply 
($\Omega_{bcc}$, $\Omega_{bbc}$, $\Omega_{bbb}$, etc.) heavy hadrons arising 
from (QCDSF-UKQCD) lattices with improved clover-Wilson light quarks. 
For the bottom quark, we use an $O(a,v^4)$-improved version of lattice 
NRQCD. 
Part of the bottomonia spectrum is used to provide an alternative scale 
and to determine the physical quark mass and radiative corrections used in 
the heavy-quark action. 
Results for spin splittings, opposite parities, and, in some cases, excited 
states are presented. 
Higher lying states and baryons with two light quarks appear to be 
especially affected by the relatively small volumes of this (initially) 
initial study. 
This and other systematics are briefly discussed.
\end{abstract}
\maketitle

\section{Introduction}
\label{SectIntroduction}

Ditto the above \cite{MakeItShort}.

Here are some references concerning experimental and 
lattice-QCD results for heavy-hadron spectroscopy: 

New bottomonia, including the ground-state $\eta_b$ 
\cite{Aubert:2008ba,Aubert:2009as,Bonvicini:2009hs} 
and the corresponding ``radial'' excitation $\eta_b(2S)$ 
\cite{Dobbs:2012zn,Mizuk:2012pb}, 
the spin-singlet $P$-waves $h_b(1P)$ and $h_b(2P)$ \cite{Adachi:2011ji}, 
and a possible $\chi_b(3P)$ \cite{Aad:2011ih}; 

Orbitally excited $B$ and $B_s$ mesons 
\cite{Abazov:2007vq,Aaltonen:2007ah,Abazov:2007sna,Aaij:2012uva}; 

$\Sigma_b^{(*)}$ \cite{Sigmab}, 
$\Xi_b$ \cite{Abazov:2007ub,Aaltonen:2007un}, and 
$\Omega_b$ baryons 
\cite{Abazov:2008qm,Aaltonen:2009ny,Aaij:2013qja}. 
Excited $\Xi_b$ \cite{Chatrchyan:2012ni} 
and $\Lambda_b$ baryons \cite{Aaij:2012da,Palni:2013bza}; 

Lattice studies of bottomonia 
\cite{Meinel:2009rd,Burch:2009az,Meinel:2010pv}, 
including those with charm sea-quarks 
\cite{Dowdall:2011wh,Dowdall:2013jqa}, 
and predictions of $D$-wave 
\cite{Daldrop:2011aa} 
and higher states 
\cite{Lewis:2012ir}; 

Lattice results for $B$ and $B_s$ mesons 
\cite{Green:2003zza,Burch:2006mb,Burch:2008qx,Detmold:2008ww,Michael:2010aa,Gregory:2010gm,McNeile:2012qf,Dowdall:2012ab}, 
including predictions for the $B_c$ system 
\cite{Allison:2004be,Gregory:2009hq}; 

Lattice results for $b$-baryons 
\cite{Lewis:2008fu,Na:2008hz,Burch:2008qx} 
and triply bottom baryons 
\cite{Meinel:2010pw,Meinel:2012qz}.

Some more recent developments: 
Lattice studies of $bc$ baryons \cite{Brown:2014ena}, 
bottomonia \cite{Wurtz:2014uca}, and 
positive parity $B_s$ mesons \cite{Lang:2015hza}; and 
the experimentalists have been busy with the $\chi_b(3P)$ state and 
$\Xi_b$ baryons \cite{Aaij:2014hla,Aaij:2014lxa,Aaij:2014yka}.

For heavy-quark-model calculations, see, e.g., 
Ref.~\cite{Eakins:2012jk}.

Once again \cite{MakeItShort}.

\section{Lattice calculation}
\label{SectLattice}

In the present section we give the details of the simulations: 
the gauge configurations used, how the quark propagators are calculated, 
and how these are put together to form the correlators for the hadrons 
of interest.

\subsection{Configurations}

We use gauge configurations which include either 
$N_f=2$ flavors of non-perturbatively improved clover-Wilson quarks 
\cite{QCDSF_Nf2} or 
$N_f=2+1$ flavors of SLiNC quarks \cite{QCDSF_Nf2p1}. 
The relevant parameters of the ensembles used can be found in 
Table \ref{latticetable}. 
As can be seen in the last column, the spatial extent of the lattices 
is rather small. 
Whereas this may not strongly affect tightly bound, multiply heavy systems 
(e.g., $\Upsilon$, $B_c$, $\Omega_{bcc}$, etc.), higher excitations and 
hadrons with lighter valence quarks may ``feel the pinch'' 
\cite{Bali:1998pi} and this is therefore a source of systematic error which 
we must keep in mind.

\begin{table}[!b]
\caption{
Relevant lattice parameters \cite{QCDSF_Nf2,QCDSF_Nf2p1}.
}
\label{latticetable}
\begin{center}
\begin{tabular}{ccccccc}
\hline \hline
lbl & $\beta$ & $\kappa_{ud}$,$\kappa_s$ & $N_s^3 \times N_t$ & $N_{conf}$ & $aM_\pi$ & $M_\pi L$  \\
\hline
{\bf a} & 5.29 & 0.13632, -- & $32^3 \times 64$ & 660 & 0.1070(5) & 3.42  \\
%{\bf a$'$} & 5.29 & 0.13632, -- & $40^3 \times 64$ & 310 & 0.1050(3) & 4.20  \\
{\bf b} & 5.50 & 0.12104, 0.12062 & $24^3 \times 48$ & 320 & 0.1406(8) & 3.37  \\
{\bf c} & 5.50 & 0.12100, 0.12070 & $24^3 \times 48$ & 180 & 0.1515(10) & 3.64  \\
{\bf d} & 5.50 & 0.12095, 0.12080 & $24^3 \times 48$ & 210 & 0.1661(8) & 3.99  \\
{\bf e} & 5.50 & 0.12090, 0.12090 & $24^3 \times 48$ & 800 & 0.1779(6) & 4.27  \\
\hline \hline
\end{tabular}
\end{center}
\end{table}

\subsection{Light and charm quark propagators}

The propagators for the light ($u,d$), strange ($s$), and charm ($c$) quarks 
were produced using the Chroma software library \cite{Chroma}. 
The light-quark propagators on ensemble {\bf a} and the light- and strange-quark 
propagators on ensembles {\bf b} and {\bf e} were originally created for other 
projects (see \cite{QCDSF_Nf2} and \cite{Bali:2011dc}, respectively) and we 
must work with the (rather severe) quark-source smearings chosen therein (see 
Table \ref{smeartable}; of course, with great computational advantage of just 
having to read in the files). 
The light- and strange-quark propagators on ensembles {\bf c} and {\bf d} and the 
charm-quark propagators on ensembles {\bf b}--{\bf e} were created with the aim of 
better resolving excited states (less smearing, see Table \ref{smeartable}). 
The charm-quark mass was taken from a related study \cite{Bali:2011dc,Bali:2012ua} 
($\kappa_c=0.1109$). The lattice scale there, however, was set using a different 
observable (the flavor-singlet baryon-mass combination $X_N$ \cite{QCDSF_Nf2p1}) 
than the one here ($M(1P)-M(1S)$ from $b\bar b$) and we therefore have a 
systematic shift in our $B_c$-system masses when using the bottomonia scale 
(see Sec.~\ref{subsect_Bmesons}).

\begin{table}[!b]
\caption{
Quark source smearings.
}
\label{smeartable}
\begin{center}
\begin{tabular}{lccr}
\hline \hline
lbl & quark & sm.type & params. \\
\hline
{\bf a} & $u,d$ & Gauss \cite{Jacobi} + & $(\kappa=0.25,N=400) +$  \\
 & & APE \cite{APE} & $(f=2.5,N=25)$  \\
{\bf b} & $u,d,s$ & Gauss + & $(\kappa=0.25,N=150) +$  \\
 & & APE & $(f=2,N=20)$  \\
{\bf c},{\bf d} & $u,d,s$ & Gauss + & $(\kappa=0.25,N=20) +$  \\
 & & APE & $(f=2,N=3)$  \\
{\bf e} & $u,d,s$ & Gauss + & $(\kappa=0.3,N=130) +$  \\
 & & APE & $(f=2,N=20)$  \\
{\bf b}--{\bf e} & $c$ & Gauss + & $(\kappa=0.25,N=12) +$  \\
 & & APE & $(f=2,N=3)$  \\
{\bf a}--{\bf e} & $Q$ & Gauss & $(\kappa=0.2,N=16)$  \\
\hline \hline
\end{tabular}
\end{center}
\end{table}

\subsection{Non-relativistic quark propagators}

For the bottom quark, we employ improved NRQCD \cite{Lepage:1992tx}, 
including terms up to $O(v^4)$, where $v$ is the heavy-quark velocity. 
We use the time-step symmetric form of the evolution equation: 
\begin{eqnarray}
  \phi(\mathbf{y},t+a) &=& 
  \left(1-\frac{a\delta H(t+a)}{2}\right)\left(1-\frac{aH_0(t+a)}{2n}\right)^n  \nonumber \\
  && \cdot U_4^\dagger(t) 
  \left(1-\frac{aH_0(t)}{2n}\right)^n\left(1-\frac{a\delta H(t)}{2}\right)  \nonumber \\
  && \cdot \phi(\mathbf{x},t) \; , 
  \label{HQevolve}
\end{eqnarray}
where the binomial expression of the exponential of the lower-order 
(in $v$) terms is carried out to $n=4$. 
$H_0$ handles the heavy-quark kinetic energy and the associated $O(a)$ 
time-step correction, 
\begin{equation}
  H_0 = -\frac{\tilde{\Delta}}{2m_Q} - \frac{a}{4n}\frac{\tilde{\Delta}^2}{4m_Q^2} \; , 
\end{equation}
and $\delta H$ contains the $O(v^4)$ relativistic corrections, 
\begin{eqnarray}
  \delta H &=& -\frac{c_1}{8m_Q^3}\tilde{\Delta}^2  \nonumber \\
  && + \frac{igc_2}{8m_Q^2}(\nabla\cdot\mathbf{E}-\mathbf{E}\cdot\nabla)  \nonumber \\
  && - \frac{gc_3}{8m_Q^2}\mathbf{\sigma}\cdot(\tilde{\nabla}\times\mathbf{E}-\mathbf{E}\times\tilde{\nabla})  \nonumber \\
  && - \frac{gc_4}{2m_Q}\mathbf{\sigma}\cdot\mathbf{B} \; . 
\end{eqnarray}
Covariant derivatives and Laplacians with a tilde represent improved versions, 
where next-to-nearest neighbor sites are included (this is the same form of 
corrections we used in previous studies of bottomonia \cite{Burch:2007fj}; the 
form of the heavy-quark evolution, Eq.~(\ref{HQevolve}), has been improved for 
the present study \cite{NRcode}). 
The (chromo) electric and magnetic fields are constructed using the standard 
four-plaquette clover formalism (see, e.g., Ref.~\cite{Burch:2003zf}) 
and tadpole improvement \cite{Lepage:1992tx,Lepage:1992xa} is applied throughout: 
$U_\mu(x) \to U_\mu(x)/u_0$, $\mathbf{E}(x) \to \mathbf{E}(x)/u_0^4$, 
$\mathbf{B}(x) \to \mathbf{B}(x)/u_0^4$, where $u_0$ is determined from the 
average plaquette. 
On each ensemble we run two or three different heavy-quark masses ($am_Q=1.5 - 3.0$) 
and for the radiative corrections, we use tree-level values $c_i=1$, as well as the 
case where $c_4=1.2$.

\subsection{Hadron correlators}
\label{subsect_corrs}

As already mentioned above (Table \ref{smeartable}), we use a combination of 
source smearings for all quark propagators. 
For the heavy quark ($Q$), we use the smeared source, as well as another where 
a covariant Laplacian is also applied (giving a radial node; for $P$- and 
$D$-wave mesons, we use a local source as the second choice). 
For all quarks we use local sinks, while for the heavy quark we also consider the 
two smearings used at the source. 
This leads to a rather peculiar situation where the heavy-quarkonia and 
triply-heavy-baryon correlators 
form $2 \times 3$ matrices, with a $2 \times 2$ symmetric block, whereas all 
correlators involving light, strange, or charm quarks, together with heavy ones, 
form $2 \times 3$ off-diagonal blocks of a larger (mostly unknown) matrix. 
This is a not a major problem, however, as we can still fit such heavy-light 
correlators to the usual ansatz, 
\begin{eqnarray}
  C(t)_{ij} &=& \langle \, 0 \, | \, O_i(t) \; O_j^\dagger(0) \, | \, 0 \, \rangle 
  \nonumber \\
  &=& \sum_{n=1}^\infty v_i^{(n)} v_j^{(n)*} \, e^{-t \, E^{(n)}} \; , 
  \label{corrhilbert}
\end{eqnarray}
except that we must fix one amplitude for each energy level considered (the fit 
then gives amplitude ratios, but the same energies). 
For some fits, we find it advantageous to consider a submatrix ($2 \times 2$ or 
$2 \times 1$) of the ones we have (this is likely due to limited statistics) 
and for most mass differences reported, we use appropriate (jackknifed) 
combinations of only the smeared-source, smeared-sink correlators (see below).

The interpolating operators that we use to combine the quarks together into the 
mesons of interest are shown in Tables \ref{QQ_op_table} and \ref{Qq_op_table}. 
One needs to be careful when combining the $u,d,s,c$ quark propagators from 
Chroma with the nonrelativistic $Q$ propagators: a change in the spin-basis is 
needed \cite{Chroma_qrot}.

\begin{table}[!b]
\caption{
Quarkonia operators.
}
\label{QQ_op_table}
 \begin{tabular}{lccc}
 \hline
 \hline
 lowest state & $J_{min}^{PC}$ & irrep $\Lambda$ & operator \\
 \hline
 $\eta_b^{}$ & $0^{-+}$ & $A_1$ & $\chi^{\dagger}\phi$ \\
 $\Upsilon$ & $1^{--}$ & $T_1$ & $\chi^{\dagger}\sigma_i\phi$ \\
 $\chi_{b0}^{}$ & $0^{++}$ & $A_1$ & $\chi^{\dagger}\sum_i\sigma_i\nabla_i\phi$ \\
 $\chi_{b1}^{}$ & $1^{++}$ & $T_1$ & $\chi^{\dagger}\sum_{jk}\epsilon_{ijk}\sigma_j\nabla_k\phi$ \\
 $\chi_{b2}^{}$ & $2^{++}$ & $T_2,E$ & $\chi^{\dagger}(\sigma_i\nabla_j+\sigma_j\nabla_i-\frac{2}{3}\delta_{ij}\sum_k\sigma_k\nabla_k)\phi$ \\
 $h_b^{}$ & $1^{+-}$ & $T_1$ & $\chi^{\dagger} \nabla_i \phi$ \\
 $\eta_{b2}^{}$ & $2^{-+}$ & $E$ & $\chi^{\dagger}(\nabla_i\nabla_i-\nabla_j\nabla_j)\phi$ \\
% $\eta_{b2}^{}$ & $2^{-+}$ & $T_2,E$ & $\chi^{\dagger}(\nabla_i\nabla_j+\nabla_j\nabla_i-\frac{2}{3}\delta_{ij}\nabla_k\nabla_k)\phi$ \\
 $\Upsilon_2$ & $2^{--}$ & $E$ & $\chi^{\dagger} ( \nabla_i\nabla_j\sigma_k + \nabla_j\nabla_k\sigma_i ) \phi$ \\
 \hline
 \hline
\end{tabular}
\end{table}

\begin{table}[!b]
\caption{
Heavy-light meson operators (for the relativistic quark: 
$q=\left( ^{u}_{l} \right)$ \cite{Chroma_qrot}).
}
\label{Qq_op_table}
 \begin{tabular}{lccc}
 \hline
 \hline
 lowest state & $J_{min}^P$ & irrep $\Lambda$ & operator \\
 \hline
 $B_q$ & $0^-$ & $A_1$ & $\chi^{\dagger} u$ \\
 $B_q^*$ & $1^-$ & $T_1$ & $\chi^{\dagger} \sigma_i u$ \\
 $B_{q0}^*$ & $0^+$ & $A_1$ & $\chi^{\dagger} l$ \\
 $B_{q1}^*$ & $1^+$ & $T_1$ & $\chi^{\dagger} \sigma_i l$ \\
 \hline
 \hline
\end{tabular}
\end{table}

The baryon operators can be found in Table \ref{Qqq_op_table}. 
In order to project out the desired spin and parity, the baryon 
correlators should then be of the form 
\begin{equation}
  B^{1/2^\pm}(t) = \left\langle \frac12(1\pm\gamma_4) O \bar O \right\rangle 
\end{equation}
or, for operators with an open Lorentz index, 
\begin{equation}
  B_{ij}^{J^\pm}(t) = \left\langle \frac12(1\pm\gamma_4) P^J_{ik} \, O_k \bar O_j 
  \right\rangle \; , 
\end{equation}
where the zero-momentum spin-projectors are 
$P^{3/2}_{ik} = \delta_{ik} - \frac13 \gamma_i \gamma_k$ and 
$P^{1/2}_{ik} = \frac13 \gamma_i \gamma_k$. 
In the end, we average over the nine remaining spatial indices ($i,j$). 
For the flavor projections, we follow lowest-order HQET and do not 
consider mixings between the different heavy-quark configurations (e.g., 
between $\Lambda_Q$ and $\Lambda^2_Q$ in Table \ref{Qqq_op_table}; 
we mostly consider the heavy-light diquark configurations in order to 
reach the negative-parity states). 
Although it is a poorer approximation, we do the same when considering 
different charm-quark configurations ($q$ or $q'=c$). 
Depending on the desired state, however, it may be that one must take 
care with the appropriate flavor projections of the light quarks 
($q,q'=u,d,s$) \cite{Oops}.

\begin{table}[!b]
\caption{
Heavy baryon operators (for the NR quark: $Q=\left( ^{\phi}_{0} \right)$ 
\cite{Chroma_qrot}).
}
\label{Qqq_op_table}
 \begin{tabular}{lcccc}
 \hline
 \hline
 label & states & $J_{min}^P$ & $s_{dq}$ & operator \\
 \hline
 $\Lambda_Q$ & $\Lambda_b,\Xi_{bc},\Omega_{bc}$ & $\frac12^+$ & 0 & $(q C \gamma_5 q') Q$ \\
 $\Lambda^2_Q$ & $\Lambda_b^{(*)},\Xi_{bc}^{('','*)},\Omega_{bc}^{('','*)}$ & $\frac12^\pm$ & 0 & $(q C \gamma_5 Q) q'$ \\
 $\Sigma_{Qi}$ & $\Sigma_b^{(*)},\Xi_b^{(',*)},\Omega_b^{(*)},\Xi_{bc}^{(',*)},\Omega_{bcc}^{(*)},...$ & $\frac12^+,\frac32^+$ & 1 & $(q C \gamma_i q') Q$ \\
 $\Sigma^2_{Qi}$ & $\Sigma_b^{(*,'*)},\Xi_b^{(',*,'*)},\Omega_b^{(*,'*)},...$ & $\frac12^\pm, \frac32^\pm$ & 1 & $(q C \gamma_i Q) q'$ \\
 $\Xi_{QQi}$ & $\Xi_{bb}^{(*,','*)},\Omega_{bb}^{(*,','*)},\Omega_{bbc}^{(*,','*)}$ & $\frac12^\pm, \frac32^\pm$ & 1 & $(Q C \gamma_i Q) q$ \\
 $\Omega_{QQQi}$ & $\Omega_{bbb}$ & $\frac32^+$ & 1 & $(Q C \gamma_i Q) Q$ \\
 \hline
 \hline
\end{tabular}
\end{table}

We also create non-zero-momentum correlators (smeared source / local sink only; 
$\vec p=2 \pi \vec n /L$, where $|\vec n| \le 3$) for the 
%$\eta_b$, $\Upsilon$, $B$, $B^*$, $B_s$, and $B_s^*$ in order to determine 
$\Upsilon$, $B^*$, and $B_s^*$ in order to determine 
the kinetic masses of these mesons from their dispersion relations. 
This provides us with an absolute mass scale and a way to set (or interpolate to) 
the physical $b$-quark mass.

For many of the heavy baryons considered herein, we present an alternative 
analysis in which we try remove most of the remaining, leading uncertainties 
in the heavy-quark parameters. 
By considering appropriate combinations of (jackknifed) average correlators, we 
subtract $E(\Upsilon)/2$ for each $b$ quark and $E(B_c)-E(\Upsilon)/2$ for each 
$c$ quark. 
Inserting experimental values for $M(\Upsilon)$ and $M(B_c)$ into these mass 
differences, we arrive at more precise, absolute estimates of the heavy baryon 
masses, albeit via ``less predictive'' means.

\section{Analysis}
\label{SectAnalysis}

In the following subsections we present our analysis of the heavy hadron 
correlators, leading to our results for the associated masses 
\cite{FitResults}.

\subsection{Quarkonia}
\label{subsect_Qonia}

\begin{figure}[t]
\begin{center}
\includegraphics*[clip,width=8cm]{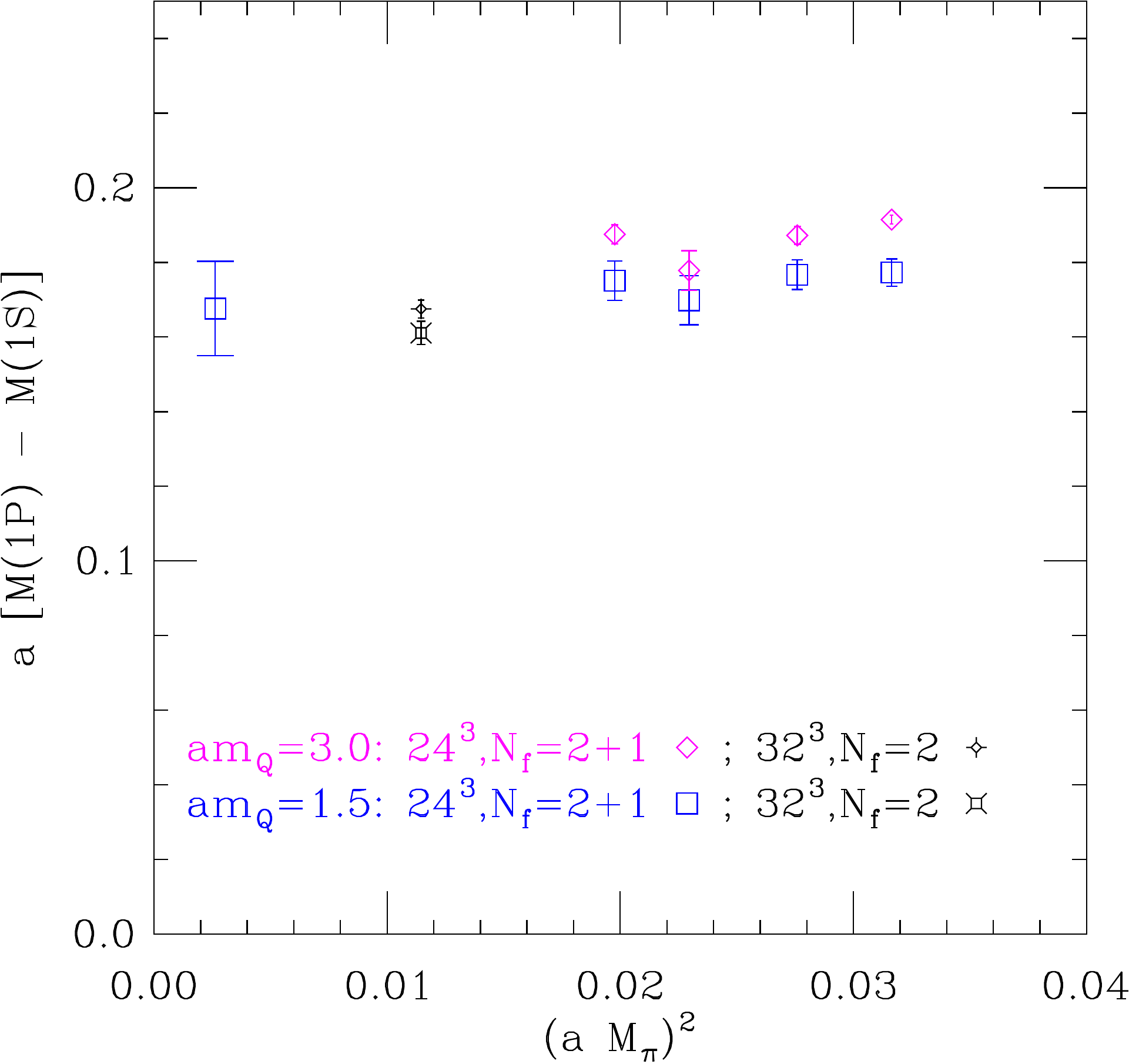}
\end{center}
\caption{
Spin-averaged $1P-1S$ $Q\bar Q$ mass differences versus pion mass.
}
\label{dM_1P-1S}
\end{figure}

At the present stage in this project we have not yet achieved a 
high-precision analysis of the bottomonia system; more precise studies 
may be found in 
Refs.~\cite{Meinel:2009rd,Burch:2009az,Meinel:2010pv,Dowdall:2011wh,
Dowdall:2013jqa,Daldrop:2011aa,Lewis:2012ir}. 
However, we need to start somewhere and, as such, we are interested in using part 
of the $b\bar b$ spectrum to set the lattice scale and the parameters used in the 
heavy-quark action ($m_Q$, $c_i$). 
We present a few other results (spin splittings and excitations) along the way. 
Further quarkonia simulations (e.g., on higher-statistics, larger-volume 
$N_f=2+1$ ensembles), hopefully leading to a more complete analysis, are 
currently underway \cite{MakeItShort}.

For spin-averaged and spin-dependent splittings we use single-elimination 
jackknife to create appropriate combinations (e.g., ratios or ratios of 
products) of the smeared-source, smeared-sink correlators to extract the 
ground-state energy differences and to handle the associated error 
correlations. 
One such example is the spin-averaged $1P-1S$ mass difference: 
\begin{eqnarray}
  \Delta M_{PS} = & [5E(\chi_{b2}) + 3E(\chi_{b1}) + E(\chi_{b0})]/9 -  \nonumber \\ 
                  & [3E(\Upsilon) + E(\eta_b)]/4 \; . 
\end{eqnarray}
It is this quantity which we use to set the scale for the lattices. 
Figure \ref{dM_1P-1S} displays the results versus the pion mass. 
The leftmost point is the chirally extrapolated $24^3$, $N_f=2+1$ difference 
(for $am_Q=1.5 \approx am_b$, see below) and the closest black point is that 
for the $32^3$, $N_f=2$ ensemble ({\bf a}). 
Using the $b\bar b$ experimental value of 457 MeV \cite{PDG}, leads to 
$a^{-1}=2726(206)$ MeV and 2837(55) MeV, respectively. 
Results from twice the heavy-quark mass ($am_Q=3.0$) are also displayed, 
showing the relatively small dependence of this splitting on $m_Q$.

\begin{figure}[t]
\begin{center}
\includegraphics*[width=8cm]{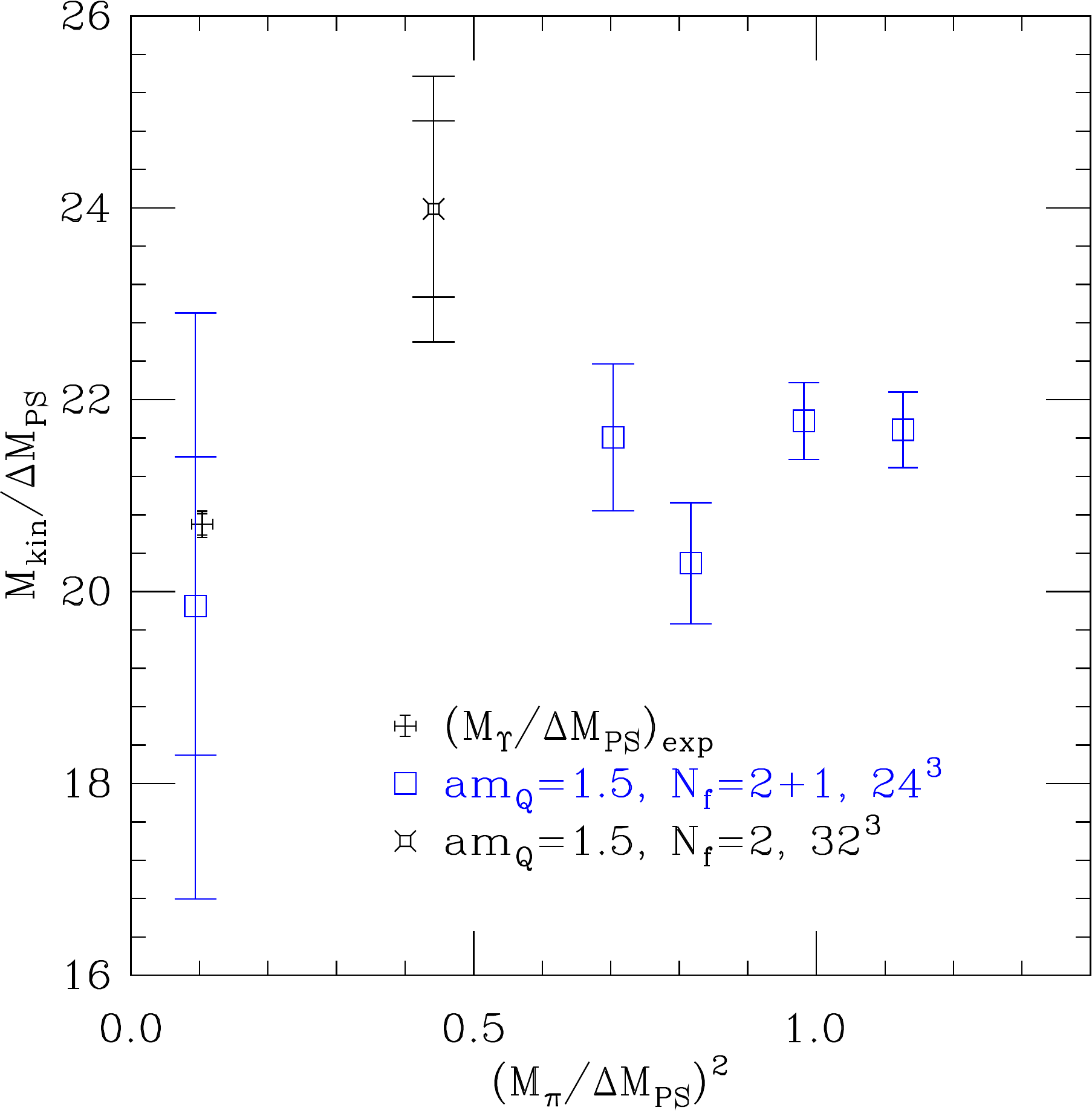}
\end{center}
\caption{
Ground-state vector $Q\bar Q$ kinetic mass versus pion mass.
}
\label{Mkin_1mm}
\end{figure}

Using ground-state energy levels from the finite-momentum $Q\bar Q$ vector 
correlators, we fit the dispersion relation to the following form: 
\begin{equation}
  E = E_0 + \frac{p^2}{2M_{kin}} - \frac{p^4}{8M_{kin}^3} \; . 
%  E = E_0 + \frac{p^2}{2M_{kin}} + A p^4 \; , 
\end{equation}
%where we check that $A$ is consistent with $-1/8M_{kin}^3$. 
Resulting values for $M_{kin}$ (in units of the $1P-1S$ splitting, 
$\Delta M_{PS}$) as a function of the pion mass are presented in 
Fig.~\ref{Mkin_1mm}. 
The larger error bars on the chirally extrapolated $24^3$ result and on the 
$32^3$ result are those which also include the error in the lattice spacing 
determination (about 7.5\% and 2\%, respectively; the same applies to all 
following figures). 
The experimental value of $M_\Upsilon/\Delta M_{PS}$ is also plotted and one 
can see agreement when $am_Q=1.5$ on the $24^3$, $N_f=2+1$ lattices. 
For the $32^3$, $N_f=2$ lattice, a slightly lower value for the heavy-quark 
mass ($am_Q \approx 1.3$) may have been appropriate. 
With the lack of a chiral extrapolation and the quenching of the strange 
quark, however, it is difficult to determine which systematics would become 
absorbed into such an adjustment. 
We use $am_Q=1.5$ as our ``working value'' of the physical bottom-quark mass 
on all ensembles.

\begin{figure}[t]
\begin{center}
\includegraphics*[width=8cm]{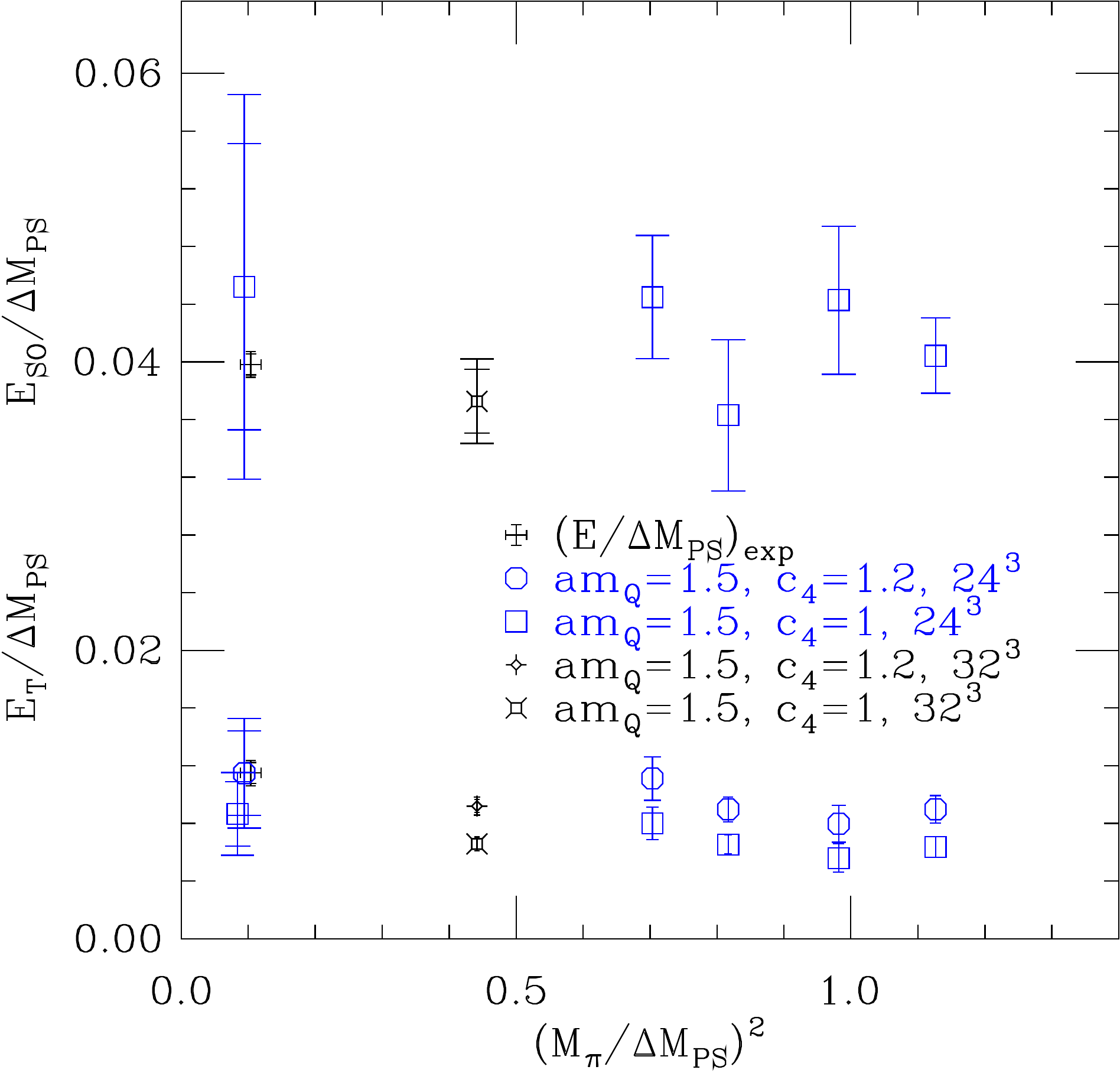}
\end{center}
\caption{
Spin-orbit (SO) and tensor (T) energies from $1P$ $Q\bar Q$ states versus pion mass, 
showing that $c_3=1$, $c_4=1.2$ works best.
}
\label{SO_T_energy}
\end{figure}

In order to find the radiative corrections for the spin-dependent terms 
($c_3$ and $c_4$) in the heavy-quark action, we look at the ``spin-orbit'' 
and ``tensor'' energies of the $1P$ $b\bar b$ system: 
\begin{equation}
  E_{SO} = \left[ -2E(\chi_{b0}) - 3E(\chi_{b1}) + 5E(\chi_{b2}) \right] / 9 \; , 
\end{equation}
\begin{equation}
  E_T = \left[ -2E(\chi_{b0}) + 3E(\chi_{b1}) - E(\chi_{b2}) \right] / 9 \; . 
\end{equation}
These are roughly proportional to $c_3$ and $c_4^2$, respectively (the 
corresponding experimental values are 18.20 MeV and 5.25 MeV). 
In Fig.~\ref{SO_T_energy}, we plot our values for these energies for the 
cases where $c_3=c_4=1$ and $c_3=1$, $c_4=1.2$. 
Within the errors the tree-level $c_3=1$ appears to work fine for $E_{SO}$, 
whereas the correction $c_4=1.2$ leads to better agreement for $E_T$. 
For the $N_f=2$ results, a slightly higher $c_4$ appears to be needed 
(at least without a chiral extrapolation); together with a lower value 
for heavy-quark mass ($am_Q=1.3$ for better $M_{kin}$ and $E_{SO}$ as well), 
the value $c_4=1.23$ would bring the tensor energy in better agreement with 
experiment. 
Now that the lattice scale and parameters of the heavy-quark action have 
been handled, we can turn our attention to other results.

\begin{figure}[t]
\begin{center}
\includegraphics*[width=8cm]{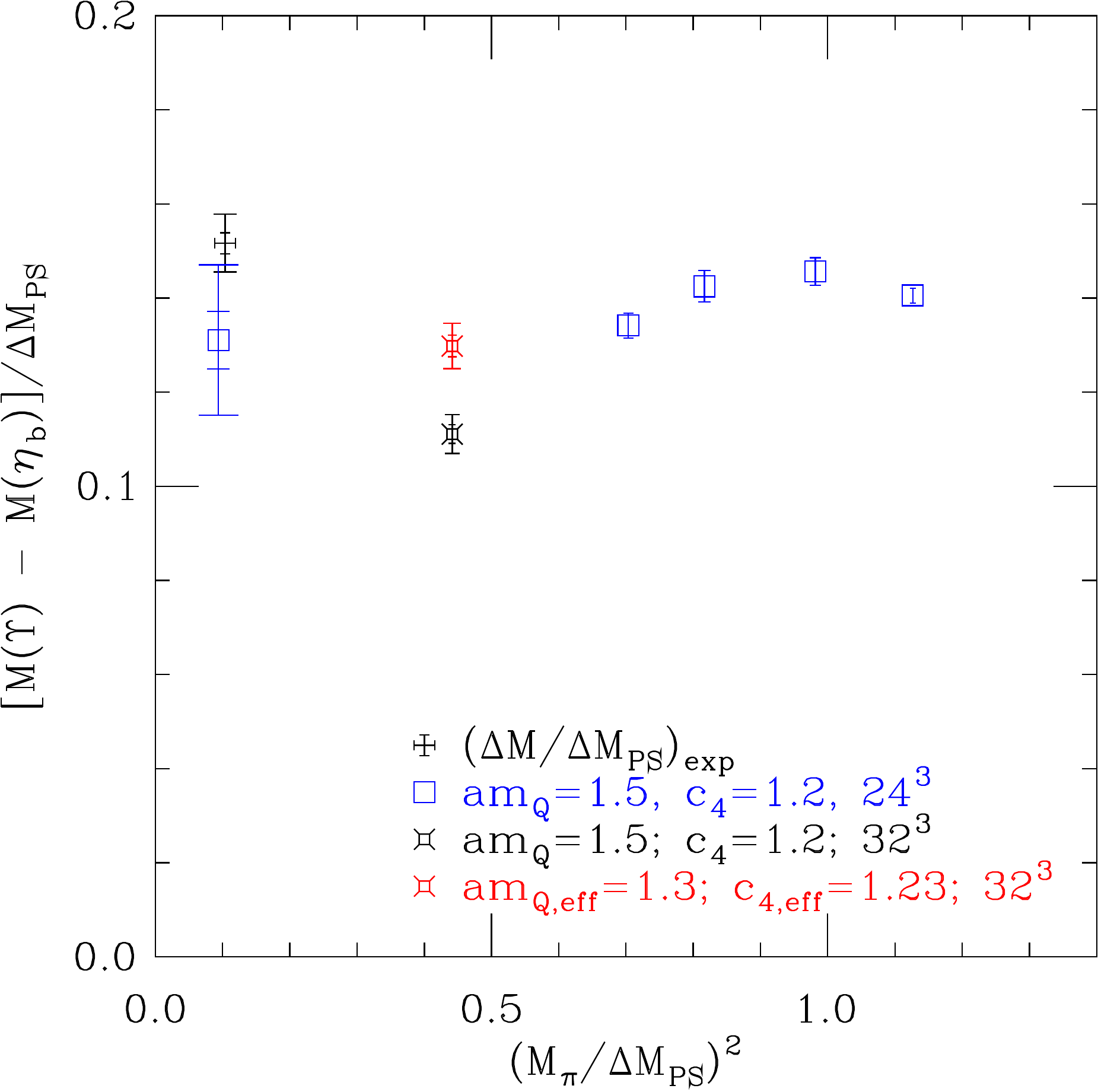}
\end{center}
\caption{
$\Upsilon - \eta_b$ mass differences versus pion mass.
}
\label{dM_1mm-0mp}
\end{figure}

Another quantity which follows from a jackknife ratio analysis is the spin 
splitting in the ground-state $b\bar b$ $S$-waves: $M(\Upsilon)-M(\eta_b^{})$. 
In Fig.~\ref{dM_1mm-0mp} one can see that the $24^3$, $N_f=2+1$ results 
extrapolate to a value slightly below the experimental value. 
However, when we include the error in the lattice spacing, we see that this 
discrepancy is only a little more than $1\sigma$. 
The $N_f=2$ result is also low, even after naively extrapolating to $am_Q=1.3$, 
$c_4=1.23$ (using the $am_Q=3.0$ and $c_4=1$ results).

We also use jackknife ratios to look more closely at the $1P$ spin splittings, 
including the $h_b$, and the energy differences to (and among) the $1D$ states. 
These results for the $24^3$, $N_f=2+1$ chiral extrapolation, the $N_f=2+1$ 
ensemble closest to the chiral limit ({\bf b}), and the $32^3$, $N_f=2$ ensemble 
({\bf a}) can be found in Table \ref{QQ_results}.

As eluded to earlier, we also fit the heavy-quarkonia correlator matrices to 
the form of Eq.~(\ref{corrhilbert}). 
Depending upon the quantum numbers and time interval being considered, we fit 
anywhere from one to three energy levels. 
This provides us with estimates of the masses of first-excited states. 
The $N_f=2$ and the chirally extrapolated $N_f=2+1$ results are compiled in 
Table \ref{QQ_results}. 
One must be careful in interpreting the results of such states. 
The continuum-spin identification in Table \ref{QQ_op_table} is the minimum 
possible ($J_{min}$) for the associated irreducible representation. 
Ideally, one should create the states of interest with operators in different 
lattice representations and look at the amplitudes as a way to confirm the 
associated spin (see, e.g., \cite{Lewis:2012ir,Thomas:2011rh}). 
For the $A_1$ operators, after 0, the next lowest possible continuum spin is 4. 
Such states should be much higher in mass and we believe that we can say with 
confidence that the first-excited states we see for such operators are still 
spin 0. 
For the other representations, the separation in possible $J$ values is not so 
large: the next lowest values are 3 for $T_1$ and $T_2$ and 4 for $E$. 
Luckily, for the case of the $T_2$ operator we have the $E$ irreducible 
representation as well and can verify that the first excitation is consistent 
for both cases (in the end we average them). 
Otherwise, since the possible continuum-spin separation for the other operators 
is at least 2 and we are only dealing with first-excited states, we assume that 
$J=J_{min}$ for these states as well.

\begin{table}[!b]
\caption{
Results for bottomonia mass splittings (in MeV) using $am_Q^{}=1.5$ and 
$c_4=1.2$ for the $N_f=2+1$ chiral limit ($\chi$), $N_f=2+1$ ensemble {\bf b}, 
and the $N_f=2$ ensemble {\bf a}. 
The first error comes from the fit, 
the second from the scale setting ($\Delta M_{PS}$).
}
\label{QQ_results}
 \begin{tabular}{lccc}
 \hline
 \hline
 splitting & $N_f=2+1$ ($\chi$) & $N_f=2+1$ ({\bf b}) & $N_f=2$ ({\bf a})  \\
 \hline
$\Upsilon - \eta_b^{}$ & 59.7(2.8)(4.5) & 58.7(1.1)(1.8) & 50.76(90)(98)  \\
$1P - \chi_{b0}^{}$ & 46.8(7.5)(3.5) & 43.9(3.1)(1.3) & 32.9(1.5)(0.6)  \\
$1P - \chi_{b1}^{}$ & 10.4(4.4)(0.8) & 9.4(1.8)(0.3) & 9.04(86)(18)  \\
$\chi_{b2}^{} - 1P$ & 15.4(4.0)(1.2) & 14.4(1.6)(0.4) & 12.01(78)(23)  \\
$1P - h_b$ & 1.3(1.8)(0.1) & 2.13(64)(6) & 1.09(50)(2)  \\
$\eta_{b2}^{} - 1S$ & 670(150)(50) & 751(58)(23) & 788(35)(15)  \\
$\Upsilon_{b2} - 1S$ & 720(190)(55) & 718(77)(22) & 824(47)(16)  \\
$\Upsilon_{b2} - \eta_{b2}^{}$ & 130(95)(10) & 60(40)(5) & 21(18)(1)  \\
 \hline
$\eta_b' - \eta_b^{}$ & 676(95)(51) & 596(31)(18) & 485(36)(9)  \\
$\Upsilon' - \Upsilon$ & 671(97)(51) & 584(33)(18) & 474(38)(9)  \\
$\chi_{b0}' - \chi_{b0}^{}$ & 560(190)(40) & 625(69)(19) & 668(93)(13)  \\  %t=7-30
$\chi_{b1}' - \chi_{b1}^{}$ & 690(200)(50) & 684(76)(21) & 729(97)(14)  \\  %t=7-30
$\chi_{b2}' - \chi_{b2}^{}$ & 510(220)(40) & 583(86)(18) & 647(78)(13)  \\  %t=7-30
$h_b' - h_b$ & 620(200)(50) & 630(79)(19) & 654(79)(13)  \\  %t=7-30
$\eta_{b2}' - \eta_{b2}^{}$ & 590(350)(45) & 810(170)(30) & 903(66)(18)  \\
$\Upsilon_{b2}' - \Upsilon_{b2}$ & 1550(400)(120) & 1170(140)(40) & 990(190)(20)  \\
 \hline
 \hline
\end{tabular}
\end{table}

\subsection{$B$, $B_s$, $B_c$ mesons}
\label{subsect_Bmesons}

Just as in the bottomonia case, we create jackknifed ratios of smeared-source, 
smeared-sink correlators to look at ground-state mass splittings of $B$ mesons. 
Table \ref{Ql_results} shows the results for the $N_f=2+1$ chiral limit, the $N_f=2+1$ 
ensemble closest to the chiral limit ({\bf b}), and the $N_f=2$ ensemble ({\bf a}).

\begin{table}[!b]
\caption{
Results for $B$-meson mass splittings (in MeV) using $am_Q^{}=1.5$ and 
$c_4=1.2$ for the $N_f=2+1$ chiral limit ($\chi$), $N_f=2+1$ ensemble {\bf b}, 
and the $N_f=2$ ensemble {\bf a}.
The first error comes from the fit, 
the second from the scale setting ($\Delta M_{PS}$).
}
\label{Ql_results}
 \begin{tabular}{lccc}
 \hline
 \hline
 splitting & $N_f=2+1$ ($\chi$) & $N_f=2+1$ ({\bf b}) & $N_f=2$ ({\bf a})  \\
 \hline
 $B^* - B$ & 45(12)(3) & 43.9(4.5)(1.3) & 49.4(3.0)(1.0)  \\
 $B_1^* - B_0^*$ & 54(11)(4) & 46.2(4.1)(1.4) & 47.3(3.2)(0.9)  \\
 $B_0^* - B$ & 281(47)(21) & 265(25)(8) & 357(29)(7)  \\  %t=7-15
 $B_1^* - B^*$ & 253(46)(19) & 256(25)(8) & 357(29)(7)  \\  %t=7-15
 \hline
 \hline
\end{tabular}
\end{table}

\begin{figure}[t]
\begin{center}
\includegraphics*[width=8cm]{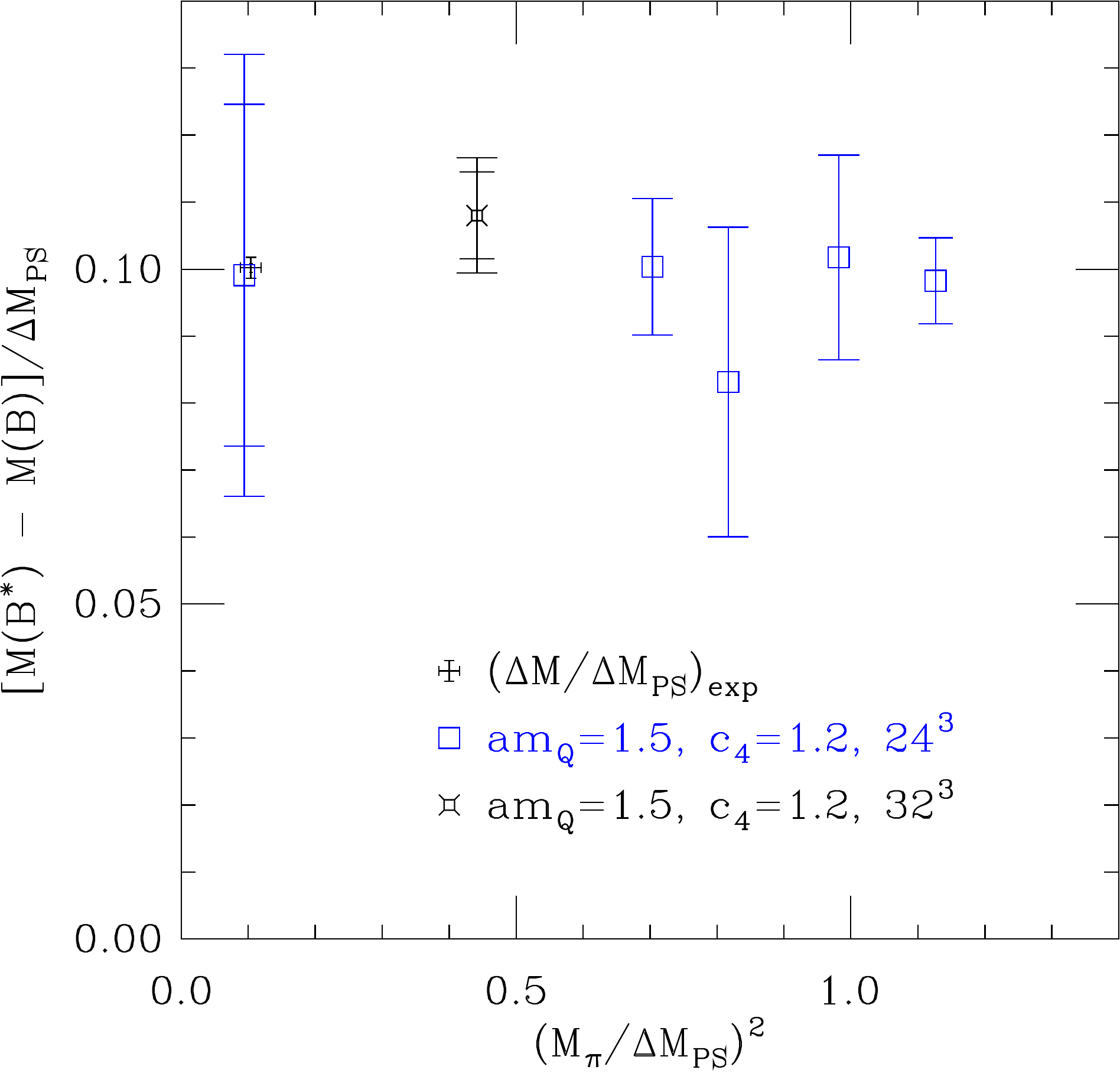}
\end{center}
\caption{
$B^*-B$ mass differences versus pion mass.
}
\label{dM_Bv-B}
\end{figure}

\begin{figure}[t]
\begin{center}
\includegraphics*[width=8cm]{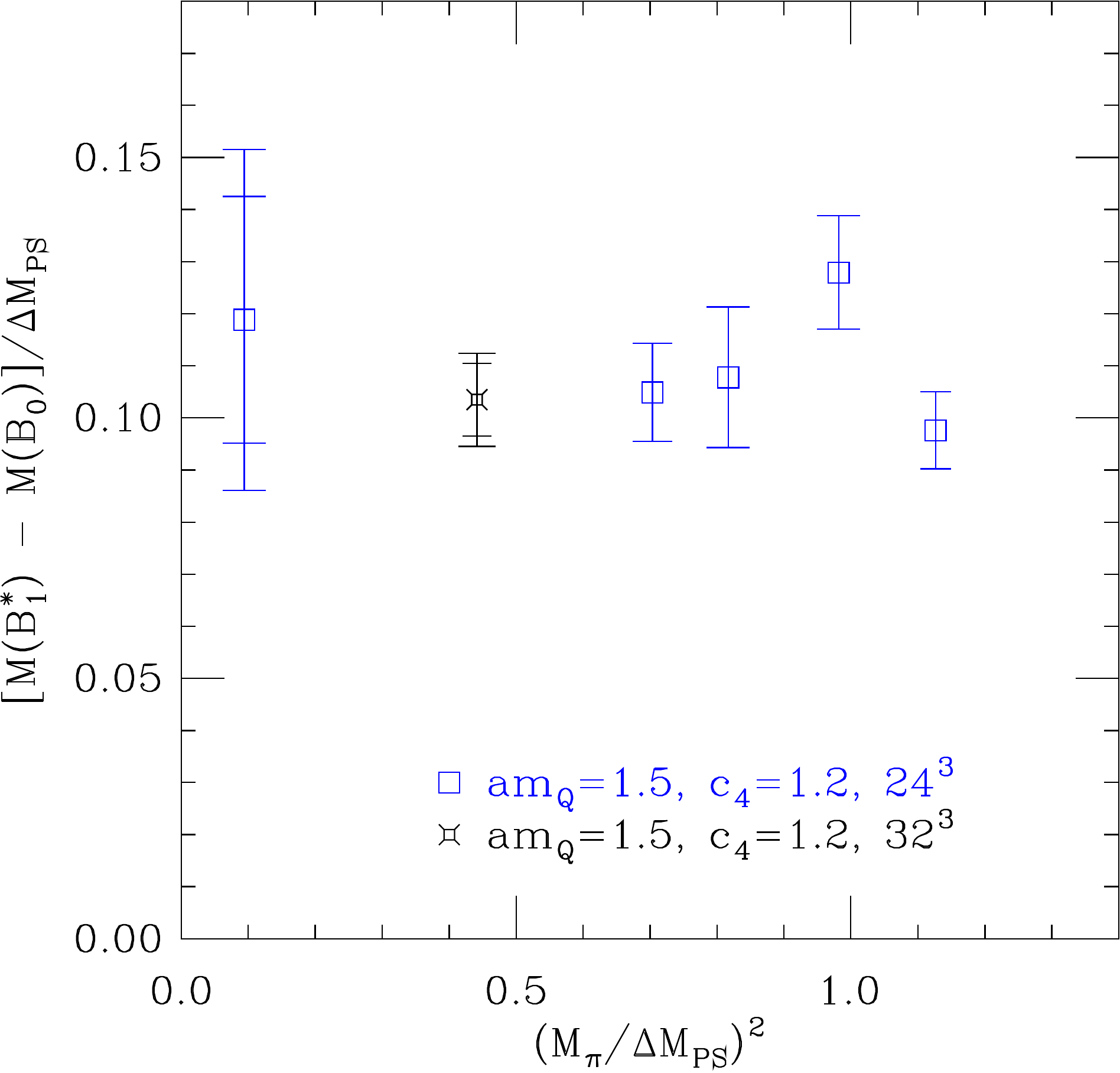}
\end{center}
\caption{
$B_1^*-B_0^*$ mass differences versus pion mass.
}
\label{dM_B1-B0}
\end{figure}

The spin splittings for the S- and P-wave $B$ mesons are shown in 
Figs.~\ref{dM_Bv-B} and \ref{dM_B1-B0} as a function of the pion mass.

On all ensembles, the $B^*-B$ difference agrees with experiment. 
Replacing the light quark with a strange or charm one leads to our results for 
$B_s$ and $B_c$ mesons. 
$N_f=2+1$ ground-state splittings, as well as differences to some first-excited 
states are shown in Table \ref{Qq_results}. 
One may note the fact that the $B^*-B$ splitting appears to be slightly 
larger than that for $B_s^*-B_s$. 
This may be a sign of the small volumes' effect on the light quarks.

\begin{table}[!b]
\caption{
Results for $B_{(s,c)}$-meson mass splittings (in MeV) using $am_Q^{}=1.5$ and 
$c_4=1.2$. for the $N_f=2+1$ chiral limit ($\chi$) and 
$N_f=2+1$ ensemble {\bf b}. 
The first error comes from the fit, 
the second from the scale setting ($\Delta M_{PS}$).
}
\label{Qq_results}
 \begin{tabular}{lcc}
 \hline
 \hline
 splitting & $N_f=2+1$ ($\chi$) & $N_f=2+1$ ({\bf b})  \\
 \hline
 $B_s^* - B_s$ & 43(10)(3) & 42.7(3.8)(1.3)  \\
 $B_{s1}^* - B_{s0}^*$ & 49.4(9.4)(3.7) & 45.4(3.4)(1.4)  \\
 $B_s^* - B^*$ & 81(21)(6) & 30.9(3.1)(0.9)  \\
 $B_{s0}^* - B$ & 418(29)(32) & 345(12)(10)  \\
 $B_{s1}^* - B^*$ & 396(29)(30) & 332(13)(10)  \\
 \hline
 $B_c^* - B_c$ & 52.0(4.5)(3.9) & 55.4(1.4)(1.7)  \\
 $B_{c1}^* - B_{c0}^*$ & 46(20)(3) & 56.5(6.5)(1.7)  \\
 $B_c^* - B^*$ & 1165(79)(88) & 1119(28)(34)  \\
 $B_{c0}^* - B$ & 1405(79)(106) & 1417(31)(43)  \\
 $B_{c1}^* - B^*$ & 1346(82)(102) & 1404(31)(43)  \\
% $B_c' - B_c$ & 810(190)(60) & 798(56)(24)  \\
% $B_c^{*'} - B_c^*$ & 790(200)(60) & 798(56)(24)  \\
 $B_c' - B_c$ & 560(260)(40) & 780(120)(25)  \\
 $B_c^{*'} - B_c^*$ & 600(250)(45) & 780(100)(25)  \\
 $B_{c0}^{*'} - B_{c0}^*$ & 460(360)(35) & 750(120)(25)  \\
 $B_{c1}^{*'} - B_{c1}^*$ & 460(310)(35) & 796(120)(25)  \\
 \hline
 \hline
\end{tabular}
\end{table}

\begin{figure}[t]
\begin{center}
\includegraphics*[width=8cm]{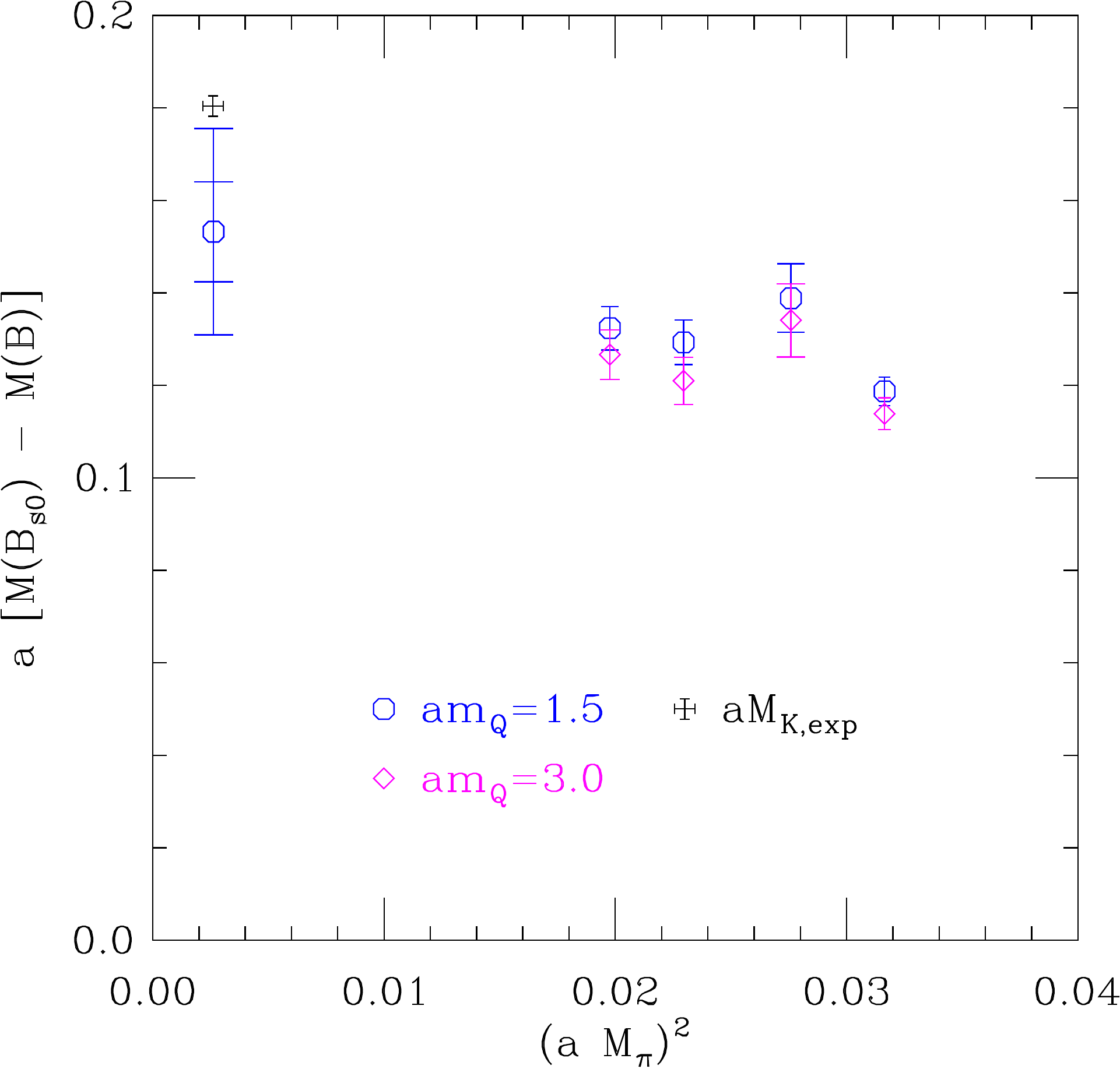}
\end{center}
\caption{
$B_{s0}^*-B$ mass differences versus pion mass. 
Indications are that the $B_{s0}^*$ is below the $BK$ threshold.
}
\label{dM_Bs0-B}
\end{figure}

Figure \ref{dM_Bs0-B} displays the $B_{s0}^*-B$ mass difference as a 
function of pion (and thereby the kaon) mass. 
All indications are that the $B_{s0}^*$ is below the $BK$ threshold. 
The same is true for the $B_{s1}^*$ and the $B^*K$ threshold (see 
Table \ref{Qq_results}).

\begin{figure}[t]
\begin{center}
\includegraphics*[width=8cm]{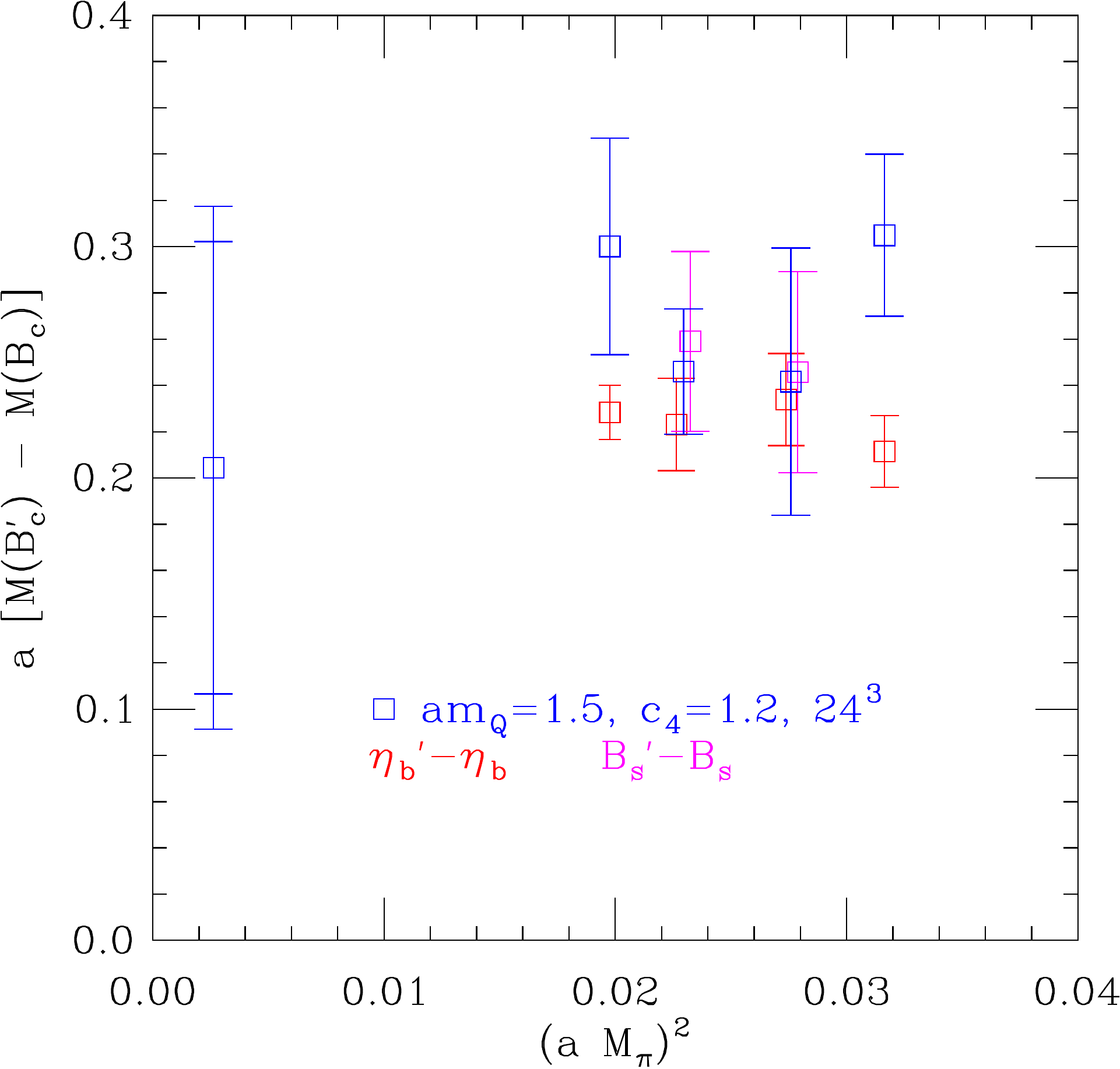}
\end{center}
\caption{
Pseudoscalar $B_c'-B_c$ mass differences versus pion mass 
($\eta_b'-\eta_b$ and $B_s'-B_s$ energies are also plotted for 
comparison).
}
\label{dM_Bcp-Bc}
\end{figure}

Due to the rather large amount of smearing used for the light- 
and strange-quark sources on ensembles {\bf b} and {\bf e} 
\cite{Bali:2011dc} (see also Table \ref{smeartable}), we are 
not able to resolve radially excited $B^{(*)}$ or $B_s^{(*)}$ mesons on 
these ensembles (and therefore, not in the chiral limit either). 
For the present study, light- and strange-quark sources with much 
less smearing were created on ensembles {\bf c} and {\bf d} to 
better study such states. 
The results for the ensemble closer to the chiral limit ({\bf c}) 
are shown in Table \ref{Qqp_results}. 
The results for the corresponding pseudoscalar $B_s'-B_s$ splitting 
appear in Fig.~\ref{dM_Bcp-Bc}, along with the $B_c'-B_c$ and 
$\eta_b'-\eta_b$ differences. 
Clearly, many more statistics are needed here, especially to reach 
a more reliable chiral limit.

\begin{table}[!b]
\caption{
Results for $B_{(s,c)}$-meson first-excited--ground state mass splittings 
(in MeV) using $am_Q^{}=1.5$ and $c_4=1.2$ on ensemble {\bf c}. 
The first error comes from the fit, 
the second from the scale setting ($\Delta M_{PS}$).
}
\label{Qqp_results}
 \begin{tabular}{lc}
 \hline
 \hline
 splitting & $N_f=2+1$ ({\bf c})  \\
 \hline
 $B' - B$ & 670(120)(30)  \\
 $B^{*'} - B^*$ & 640(90)(25)  \\
 $B_0^{*'} - B_0^*$ & 790(160)(30)  \\
 $B_1^{*'} - B_1^*$ & 760(110)(30)  \\
 \hline
 $B_s' - B_s$ & 697(100)(27)  \\
 $B_s^{*'} - B_s^*$ & 659(80)(25)  \\
 $B_{s0}^{*'} - B_{s0}^*$ & 730(110)(30)  \\
 $B_{s1}^{*'} - B_{s1}^*$ & 730(95)(30)  \\
 \hline
 $B_c' - B_c$ & 661(73)(25)  \\
 $B_c^{*'} - B_c^*$ & 662(73)(25)  \\
 $B_{c0}^{*'} - B_{c0}^*$ & 630(110)(25)  \\
 $B_{c1}^{*'} - B_{c1}^*$ & 675(95)(26)  \\
 \hline
 \hline
\end{tabular}
\end{table}

\subsection{Singly heavy baryons}
\label{subsect_b_baryons}

The naming conventions for heavy baryons may not be strictly obeyed 
here, especially where negative-parity states are concerned. 
Therefore, for the purpose of notational clarity, we briefly point out 
the quantum numbers and the corresponding names of the baryon states 
discussed in this subsection: 

$J^P(s_{dq})=\frac12^+(0)$ : $\Lambda_b$ , $\Xi_b$ 

$J^P(s_{dq})=\frac12^+(1)$ : $\Sigma_b$ , $\Xi_b'$ , $\Omega_b$ 

$J^P(s_{dq})=\frac32^+(1)$ : $\Sigma_b^*$ , $\Xi_b^*$ , $\Omega_b^*$ 

$J^P(s_{dq})=\frac12^-(0)$ : $\Lambda_b^*$ , $\Xi_b''$ 

$J^P(s_{dq})=\frac32^-(1)$ : $\Sigma_b'^*$ , $\Xi_b'^*$ , $\Omega_b'^*$ 

\noindent
where $s_{dq}$ is the spin of the diquark appearing in the associated 
baryon interpolator ({\it not} that of the state).

Table \ref{Qqq_results} lists the results for our mass splittings 
among the singly heavy baryons.

\begin{figure}[t]
\begin{center}
\includegraphics*[width=8cm]{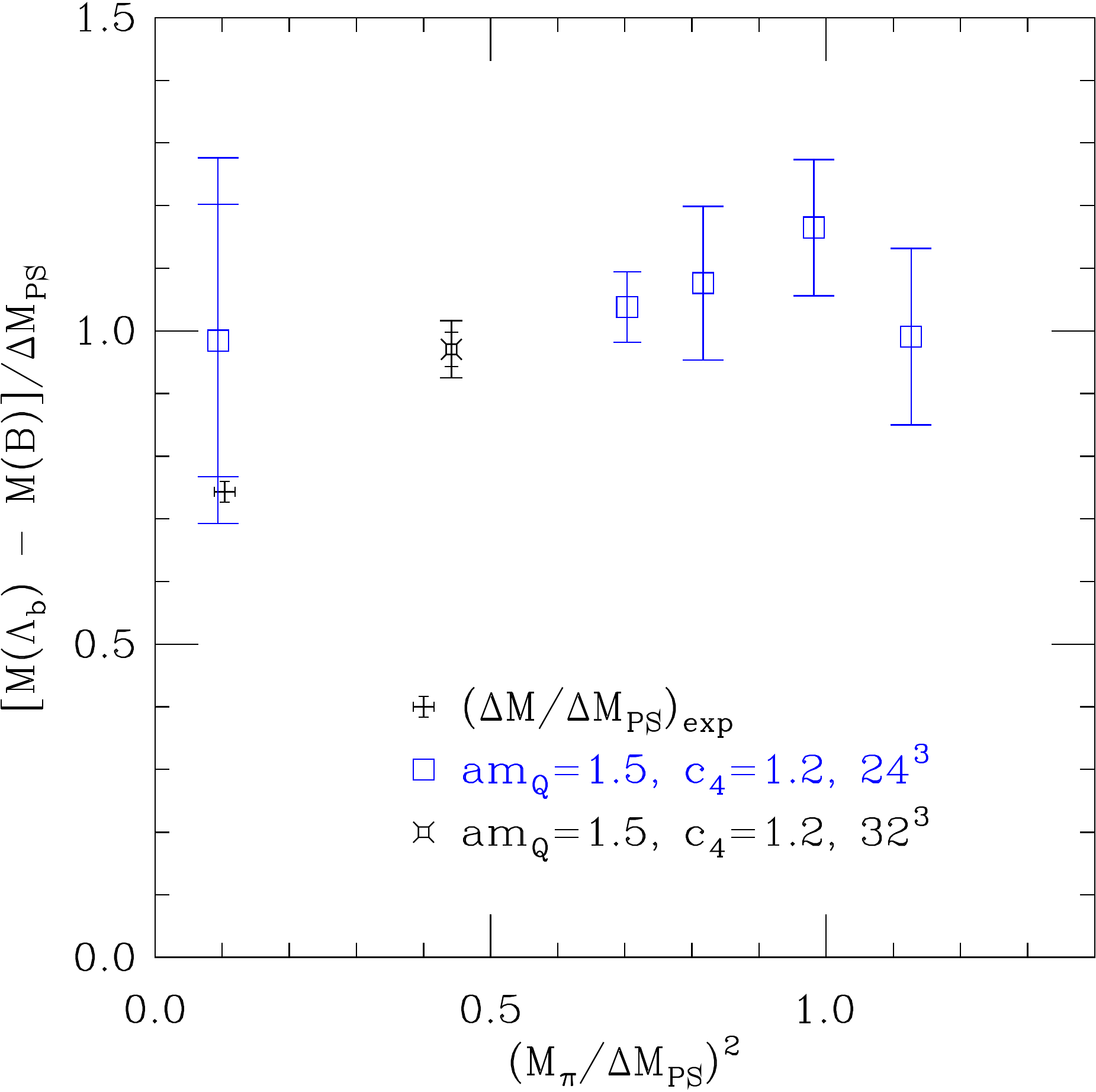}
\end{center}
\caption{
$\Lambda_b - B$ mass differences versus $M_\pi^2$.
}
\label{dM_L-B}
\end{figure}

\begin{figure}[t]
\begin{center}
\includegraphics*[width=8cm]{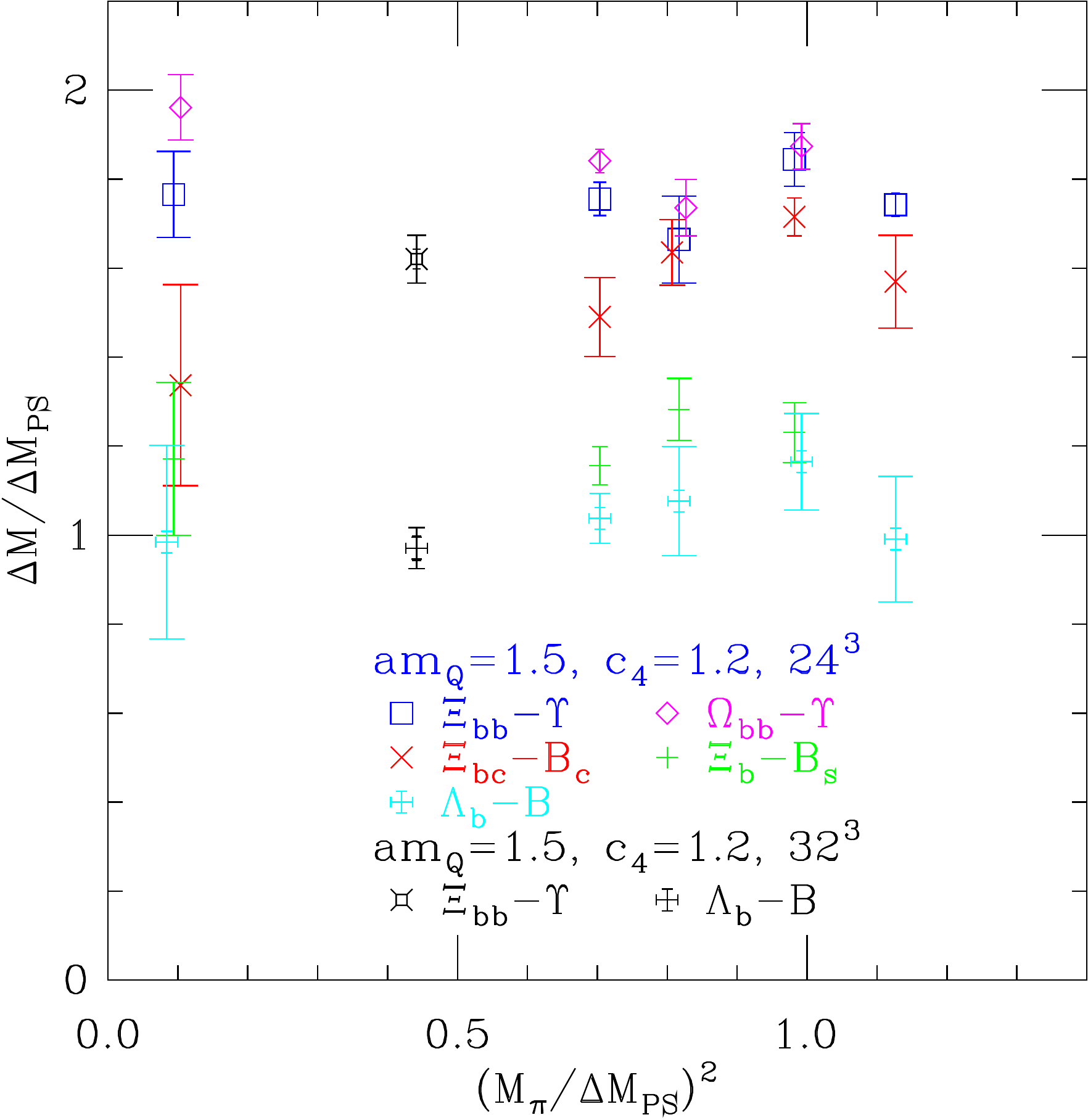}
\end{center}
\caption{
$\Xi_{bb} - \Upsilon$, $\Xi_{bc} - B_c$, $\Xi_b - B_s$, and 
$\Lambda_b - B$ mass differences versus $M_\pi^2$.
}
\label{dM_Xbb_Xbc_Xb_Lb}
\end{figure}

Figure \ref{dM_L-B} shows the $\Lambda_b - B$ mass difference as a function 
of $M_\pi^2$. 
The results on all lattices are high when compared to experiment, a sign that 
the small volumes (see Table \ref{latticetable}) may be drastically affecting 
our results for baryons with two light quarks. 
The $\Lambda_b - B$ splitting appears again in Fig.~\ref{dM_Xbb_Xbc_Xb_Lb}, 
along with the the analogous $\Xi_b - B_s$ splitting, but here again our 
lattice results are too high, 536(78) MeV ($\chi$-extrap. $N_f=2+1$), when 
compared to experiment (427 MeV). 
The $\Omega_b - B_s$ difference shows much better agreement.

\begin{figure}[t]
\begin{center}
\includegraphics*[width=8cm]{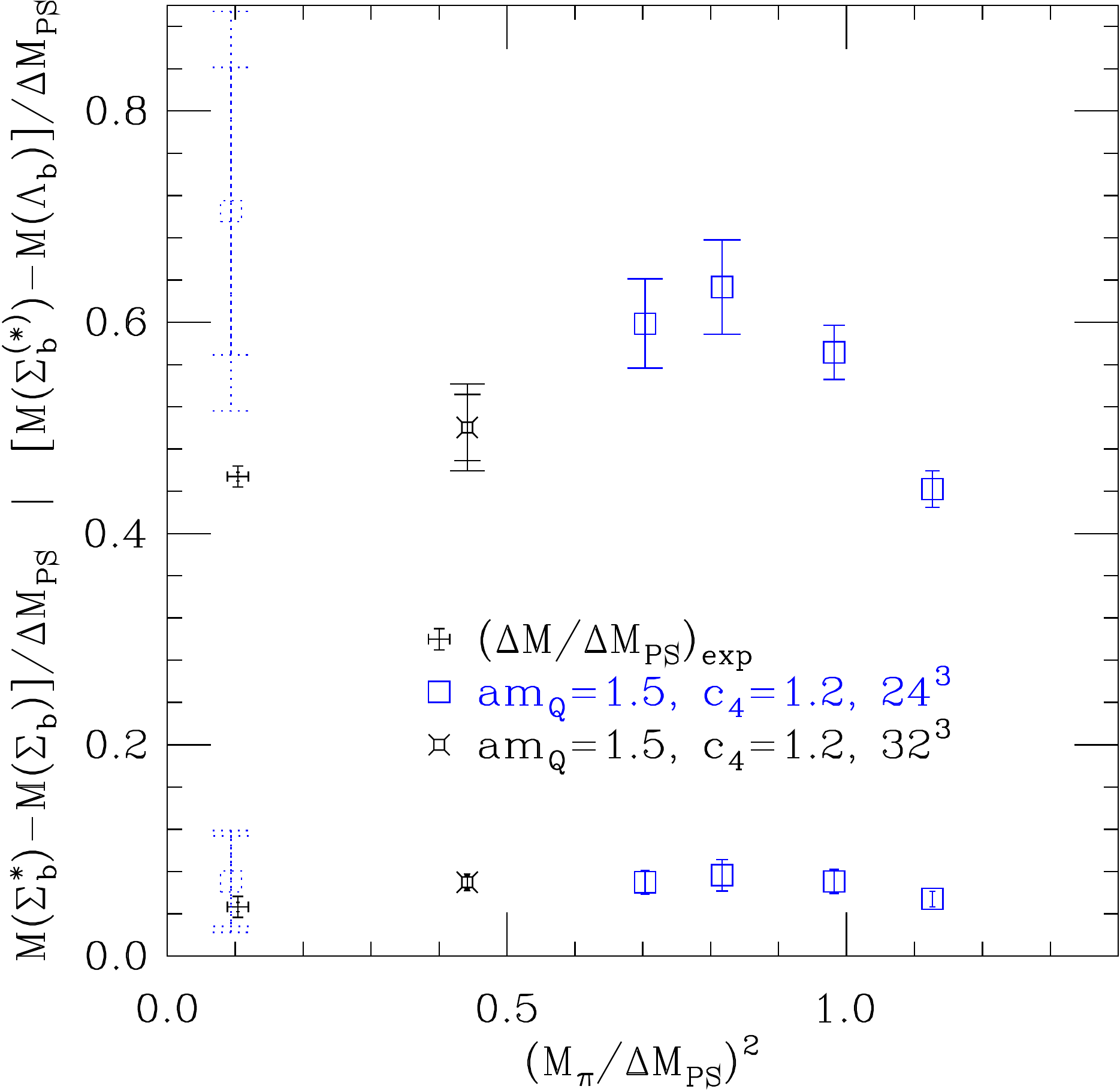}
\end{center}
\caption{
$\Sigma_b^{(*)} - \Lambda_b$ and $\Sigma_b^* - \Sigma_b$ 
mass differences versus $M_\pi^2$.
}
\label{dM_SsS-L}
\end{figure}

Mass differences from alternative spin-flavor combinations appear in 
Fig.~\ref{dM_SsS-L}. 
Here, the (spin averaged) $\Sigma_b^{(*)} - \Lambda_b$ and 
$\Sigma_b^* - \Sigma_b$ splittings also show possible signs of 
finite-volume-induced enhancements (the dotted results are from 
chiral extrapolations without the heaviest pion mass, {\bf e}). 
Other spin and multiplet splittings can be found in 
Table \ref{Qqq_results} (sometimes also with the same limited chiral 
extrapolation).

\begin{table}[!b]
\caption{
Results for $bqq'$ (where $q,q'=u,d,s$) baryon mass splittings (in MeV) 
using $am_Q^{}=1.5$ and $c_4=1.2$ for 
the $N_f=2+1$ chiral limit ($\chi$), 
the $N_f=2+1$ ensemble {\bf b}, 
and the $N_f=2$ ensemble {\bf a}. 
The first error comes from the fit, 
the second from the scale setting ($\Delta M_{PS}$). 
($^\dagger$Using only the three lightest $M_\pi$ values.)
}
\label{Qqq_results}
 \begin{tabular}{lccc}
 \hline
 \hline
 splitting & $N_f=2+1$ ($\chi$) & $N_f=2+1$ ({\bf b}) & $N_f=2$ ({\bf a}) \\
 \hline
$\Lambda_b - B$ & 450(99)(34) & 474(26)(14) & 444(12)(9) \\
$\Xi_b - B_s$ & 536(78)(40) & 528(20)(16) & -- \\
$\Omega_b - B_s$ & 746(43)(56) & 695(16)(21) & -- \\
 \hline
$\Xi_b - \Lambda_b$ & 231(26)(17) & 87.2(8.2)(2.6) & -- \\
%$\Xi_b' - \Lambda_b$ &  &  & -- \\
%$\Omega_b - \Lambda_b$ &  &  & -- \\
 \hline
$\Sigma_b^{(*)} - \Lambda_b$ & 322(62)(24)$^\dagger$ & 274(19)(8) & 229(14)(4) \\
$\Xi_b' - \Xi_b$ & 238(25)(18) & 190(10)(6) & -- \\
 & 154(40)(12)$^\dagger$ &  &  \\
$\Omega_b - \Xi_b$ & 314(34)(24) & 221(12)(7) & -- \\
 \hline
$\Sigma_b^{*} - \Sigma_b$ & 32(19)(2)$^\dagger$ & 31.9(5.1)(1.0) & 31.9(3.0)(0.6) \\
$\Xi_b^* - \Xi_b'$ & 36.6(9.8)(2.8) & 33.4(3.1)(1.0) & -- \\
$\Omega_b^{*} - \Omega_b$ & 31.6(8.5)(2.4) & 27.6(3.2)(0.8) & -- \\
 \hline
$\Lambda_b^* - \Lambda_b$ & 514(67)(39) & 455(26)(14) & 495(27)(10) \\
$\Xi_b'' - \Xi_b$ & 385(79)(29)$^\dagger$ & 398(20)(12) & -- \\
$\Sigma_b'^* - \Sigma_b^*$ & 142(30)(11) & 183(12)(6) & 274(15)(5) \\
$\Sigma_b'^* - \Lambda_b^*$ & 53(14)(4) & 46.1(6.4)(1.4) & 36.0(4.9)(0.7) \\
$\Xi_b'^* - \Xi_b^*$ & 262(36)(20) & 237(15)(7) & -- \\
$\Omega_b'^* - \Omega_b^*$ & 308(28)(23) & 249(11)(8) & -- \\
 \hline
 \hline
\end{tabular}
\end{table}

\begin{figure}[t]
\begin{center}
\includegraphics*[width=8cm]{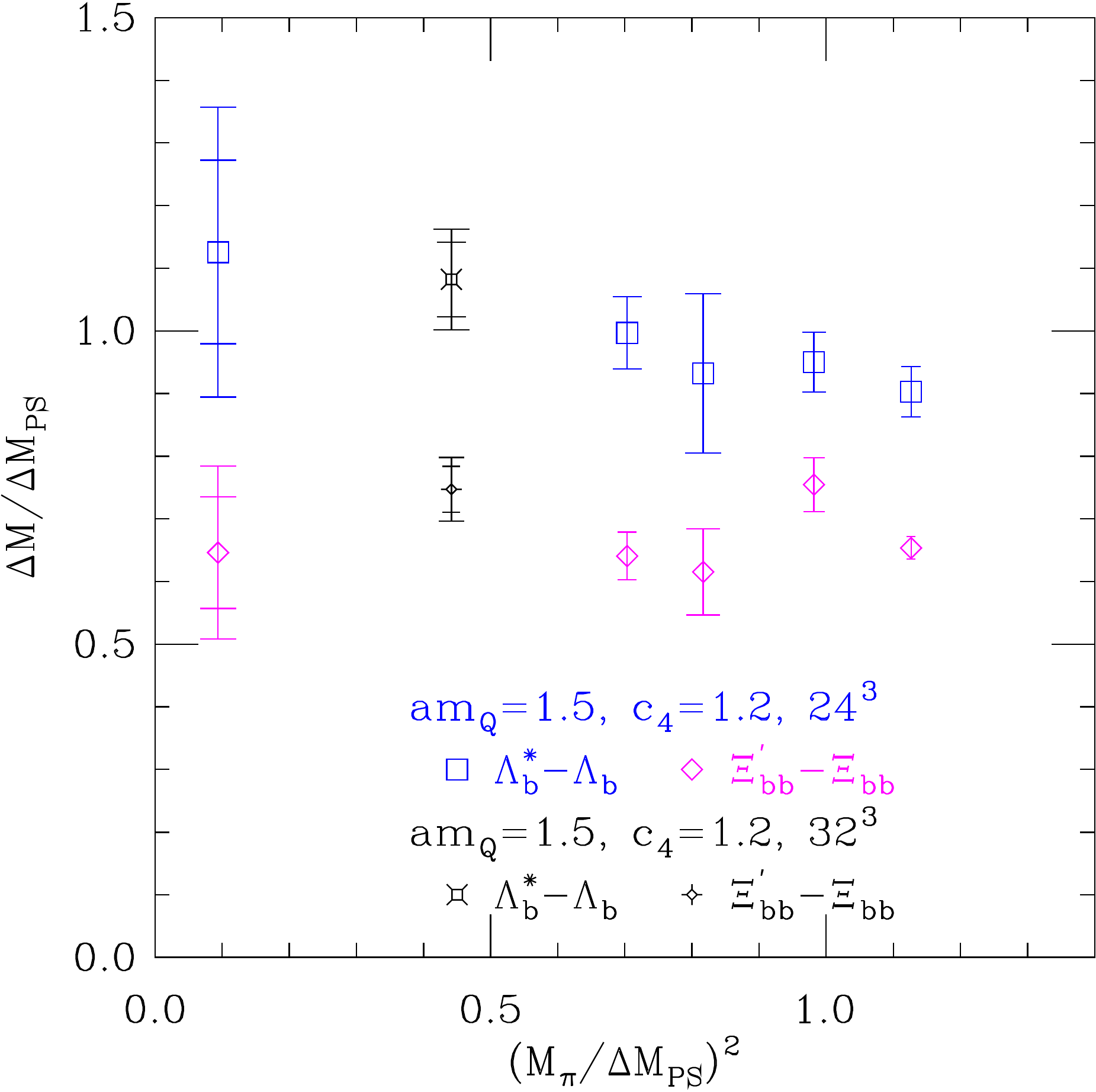}
\end{center}
\caption{
Mass differences -- involving the $\Lambda_b^*$ and 
$\Xi^{'}_{bb}$ negative-parity states -- versus $M_\pi^2$.
}
\label{OppParBar}
\end{figure}

Figure \ref{OppParBar} displays the mass differences between 
the $\Lambda_b$ and its parity partner, $\Lambda_b^*$. 
Again, there is a clear overestimate here; the experimental value being 
around 300 MeV \cite{Aaij:2012da}. 
Mass differences involving this and other negative-parity states can be 
found in Table \ref{Qqq_results} \cite{MakeItShort}.

\subsection{Doubly heavy baryons}
\label{subsect_bb_baryons}

As before, we point out the names of the states discussed in this 
subsection: 

$J^P(s_{dq})=\frac12^+(0)$ : $\Xi_{bc}$ , $\Omega_{bc}$ 

$J^P(s_{dq})=\frac12^+(1)$ : $\Xi_{bc}'$ , $\Omega_{bc}'$ , 
$\Xi_{bb}$ , $\Omega_{bb}$ 

$J^P(s_{dq})=\frac32^+(1)$ : $\Xi_{bc}^*$ , $\Omega_{bc}^*$ , 
$\Xi_{bb}^*$ , $\Omega_{bb}^*$ 

$J^P(s_{dq})=\frac12^-(0)$ : $\Xi_{bc}''$ , $\Omega_{bc}''$ 

$J^P(s_{dq})=\frac12^-(1)$ : $\Xi_{bb}'$ , $\Omega_{bb}'$ 

$J^P(s_{dq})=\frac32^-(1)$ : $\Xi_{bc}'^*$ , $\Omega_{bc}'^*$ , 
$\Xi_{bb}'^*$ , $\Omega_{bb}'^*$ 

\noindent
and any ``radial'' excitations are denoted by a (2) after the name.

Mass differences involving the $bc$-baryons can be found in 
Table \ref{Qcq_results}. 
Those for doubly bottom baryons appear in Table \ref{QQq_results}.

Also present in the previously mentioned Fig.~\ref{dM_Xbb_Xbc_Xb_Lb} 
are the results for the (yet to be observed) $\Xi_{bc} - B_c$ mass 
difference. 
Using the physical value for $M(B_c)$, we find: 

\noindent
$M(\Xi_{bc})=6887(103)(46)$ MeV ($\chi$-extrap.\ $N_f=2+1$).

\begin{table}[!b]
\caption{
Results for $bcq$ (where $q=u,d,s$) baryon mass splittings (in MeV) 
using $am_Q^{}=1.5$ and $c_4=1.2$ for 
the $N_f=2+1$ chiral limit ($\chi$) and 
the $N_f=2+1$ ensemble {\bf b}. 
The first error comes from the fit, 
the second from the scale setting ($\Delta M_{PS}$).
}
\label{Qcq_results}
 \begin{tabular}{lcc}
 \hline
 \hline
 splitting & $N_f=2+1$ ($\chi$) & $N_f=2+1$ ({\bf b}) \\
 \hline
$\Xi_{bc} - B_c$ & 611(103)(46) & 681(41)(21) \\
%$\Omega_{bc} - B_c$ &  &  \\
 \hline
$\Omega_{bc} - \Xi_{bc}$ & 152(34)(11) & 51(13)(2) \\
 \hline
$\Xi_{bc}' - \Xi_{bc}$ & 50(19)(4) & 47.2(7.2)(1.4) \\
$\Omega_{bc}' - \Omega_{bc}$ & 38(15)(3) & 43.5(4.9)(1.3) \\
 \hline
$\Xi_{bc}^* - \Xi_{bc}'$ & 26(13)(2) & 26.7(4.6)(0.8) \\
$\Omega_{bc}^* - \Omega_{bc}'$ & 21(11)(2) & 25.0(3.4)(0.8) \\
 \hline
$\Xi_{bc}'' - \Xi_{bc}$ & 290(66)(22) & 314(21)(10) \\
$\Omega_{bc}'' - \Omega_{bc}$ & 342(53)(26) & 334(14)(10) \\
 \hline
%$\Xi_{bc}'^* - \Xi_{bc}$ &  &  \\
%$\Xi_{bc}'^* - \Xi_{bc}^*$ &  &  \\
$\Xi_{bc}'^* - \Xi_{bc}''$ & 57(17)(4) & 48.1(5.4)(1.5) \\
%$\Omega_{bc}'^* - \Omega_{bc}^*$ &  &  \\
$\Omega_{bc}'^* - \Omega_{bc}''$ & 52(14)(4) & 45.6(4.0)(1.4) \\
 \hline
 \hline
\end{tabular}
\end{table}

The $\Xi_{bb}-\Upsilon$ and $\Omega_{bb}-\Upsilon$ energies may also 
be seen in Fig.~\ref{dM_Xbb_Xbc_Xb_Lb}. 
With the physical value for $M(\Upsilon)$ as input, we find: 

\noindent
$M(\Xi_{bb}) = 10201(10)(14)$ MeV ($N_f=2$) ; 

\noindent
$M(\Xi_{bb}) = 10267(44)(61)$ MeV ($\chi$-extrap.\ $N_f=2+1$) ; 

\noindent
$M(\Omega_{bb}) = 10356(34)(68)$ MeV ($\chi$-extrap.\ $N_f=2+1$).

Along with the $\Lambda_b^*$, the negative-parity $\Xi_{bb}'$ also 
appears in Fig.~\ref{OppParBar}.

For more results involving spin splittings, negative-parity states, 
and (for the $bb$-baryons) first-excited states, see 
Tables \ref{Qcq_results} and \ref{QQq_results} \cite{MakeItShort}.

\begin{table}[!b]
\caption{
Results for $bbq$ (where $q=u,d,s$) baryon mass splittings (in MeV) 
using $am_Q^{}=1.5$ and $c_4=1.2$ for 
the $N_f=2+1$ chiral limit ($\chi$), 
the $N_f=2+1$ ensemble {\bf b}, 
and the $N_f=2$ ensemble {\bf a}. 
The first error comes from the fit, 
the second from the scale setting ($\Delta M_{PS}$).
($^\dagger$Using only the three lightest $M_\pi$ values.)
}
\label{QQq_results}
 \begin{tabular}{lccc}
 \hline
 \hline
 splitting & $N_f=2+1$ ($\chi$) & $N_f=2+1$ ({\bf b}) & $N_f=2$ ({\bf a}) \\
 \hline
$\Xi_{bb} - \Upsilon$ & 807(44)(61) & 802(17)(24) & 741(10)(14) \\
$\Omega_{bb} - \Upsilon$ & 896(34)(68) & 841(12)(25) & -- \\
 \hline
$\Xi_{bb}^* - \Xi_{bb}$ & 49(12)(4) & 33.7(4.9)(1.0) & 35.1(6.5)(0.7) \\
  & 34(20)(3)$^\dagger$ &  &  \\
$\Omega_{bb}^* - \Omega_{bb}$ & 45.1(9.5)(3.4) & 32.5(3.7)(1.0) & -- \\
  & 31(17)(2)$^\dagger$ &  &  \\
 \hline
$\Xi_{bb}' - \Xi_{bb}$ & 295(41)(22) & 293(17)(9) & 341(17)(7) \\
$\Omega_{bb}' - \Omega_{bb}$ & 362(44)(27) & 299(19)(9) & -- \\
 \hline
$\Xi_{bb}'^* - \Xi_{bb}'$ & 54(12)(4) & 40.5(6.9)(1.2) & 28.4(3.4)(0.6) \\
  & 40(20)(3)$^\dagger$ &  &  \\
$\Omega_{bb}'^* - \Omega_{bb}'$ & 50(10)(4) & 37.3(5.0)(1.1) & -- \\
  & 37(17)(3)$^\dagger$ &  &  \\
 \hline
%$\Xi_{bb}(2) - \Xi_{bb}$ & -- & 671(69)(20) & 422(59)(8) \\
$\Xi_{bb}(2) - \Xi_{bb}$ & -- & 671(69)(20) & 375(110)(7) \\
$\Xi_{bb}^*(2) - \Xi_{bb}^*$ & -- & 591(63)(18) & 387(83)(7) \\
$\Omega_{bb}(2) - \Omega_{bb}$ & -- & 624(54)(19) & -- \\
$\Omega_{bb}^*(2) - \Omega_{bb}^*$ & -- & 564(50)(17) & -- \\
$\Xi_{bb}'(2) - \Xi_{bb}'$ & -- & 739(92)(22) & -- \\
$\Xi_{bb}'^*(2) - \Xi_{bb}'^*$ & -- & 793(95)(24) & -- \\
$\Omega_{bb}'(2) - \Omega_{bb}'$ & -- & 722(74)(22) & -- \\
$\Omega_{bb}'^*(2) - \Omega_{bb}'^*$ & -- & 711(71)(22) & -- \\
 \hline
 \hline
\end{tabular}
\end{table}

\subsection{Triply heavy baryons}
\label{subsect_bbb_baryons}

As before, we point out the names of the states discussed in this 
subsection: 

$J^P(s_{dq})=\frac12^+(1)$ : $\Omega_{bcc}$ , $\Omega_{bbc}$ 

$J^P(s_{dq})=\frac32^+(1)$ : $\Omega_{bcc}^*$ , $\Omega_{bbc}^*$ , $\Omega_{bbb}$ 

$J^P(s_{dq})=\frac12^-(1)$ : $\Omega_{bbc}'$ 

$J^P(s_{dq})=\frac32^-(1)$ : $\Omega_{bcc}'^*$ , $\Omega_{bbc}'^*$ 

\noindent
and any ``radial'' excitations are denoted by a (2) after the name.

Mass differences involving $bcc$ and $bbc$ baryons can be found in 
Table \ref{QQc_results}. 
Triply bottom results appear in Table \ref{QQQ_results}.

With the physical values for $M(B_c)$ and $M(\Upsilon)$ as input, we find: 

\noindent
$M(\Omega_{bcc}) = 7984(27)(12)$ MeV ($\chi$-extrap.\ $N_f=2+1$) ; 

\noindent
$M(\Omega_{bbc}) = 11182(27)(13)$ MeV ($\chi$-extrap.\ $N_f=2+1$).

\begin{figure}[t]
\begin{center}
\includegraphics*[width=8cm]{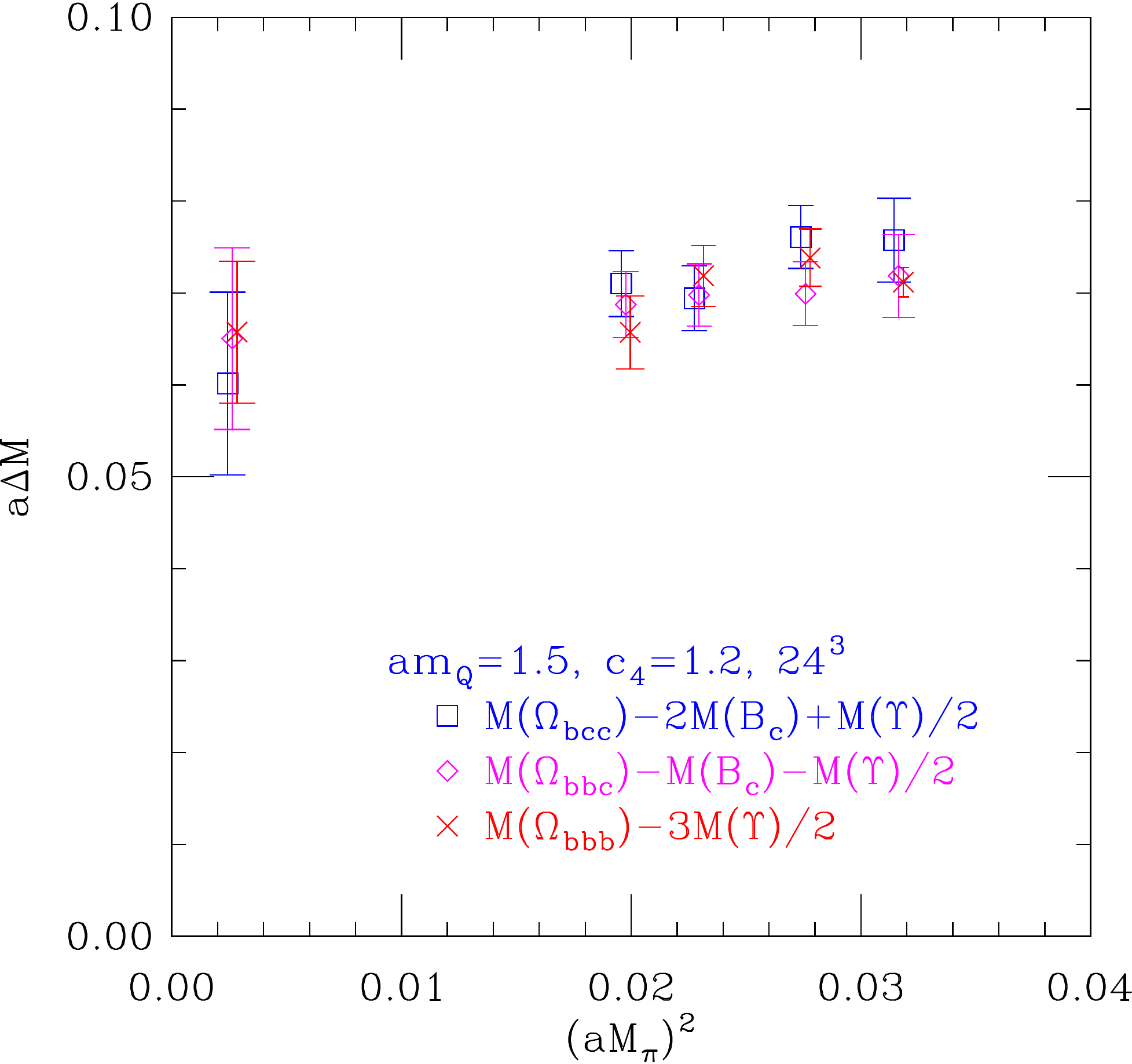}
\end{center}
\caption{
$\Omega_{bcc}$, $\Omega_{bbc}$, and $\Omega_{bbb}$ masses versus 
$M_\pi^2$.
}
\label{dM_Obxx}
\end{figure}

With the physical value for $M(\Upsilon)$ as input, we find: 

\noindent
$M(\Omega_{bbb}) = 14357.6(4.4)(3.3)$ MeV ($N_f=2$) ; 

\noindent
$M(\Omega_{bbb}) = 14369(21)(14)$ MeV ($\chi$-extrap.\ $N_f=2+1$).

Figure \ref{dM_Obxx} shows the $\Omega_{bcc}$, $\Omega_{bbc}$, and 
$\Omega_{bbb}$ masses (referenced from an appropriate combination 
of $M(\Upsilon)$ and $M(B_c)$) versus the pion mass. 
Once the heavy-quark masses are ``subtracted'', seemingly little 
difference remains between these triply heavy systems.

\begin{figure}[t]
\begin{center}
\includegraphics*[width=8cm]{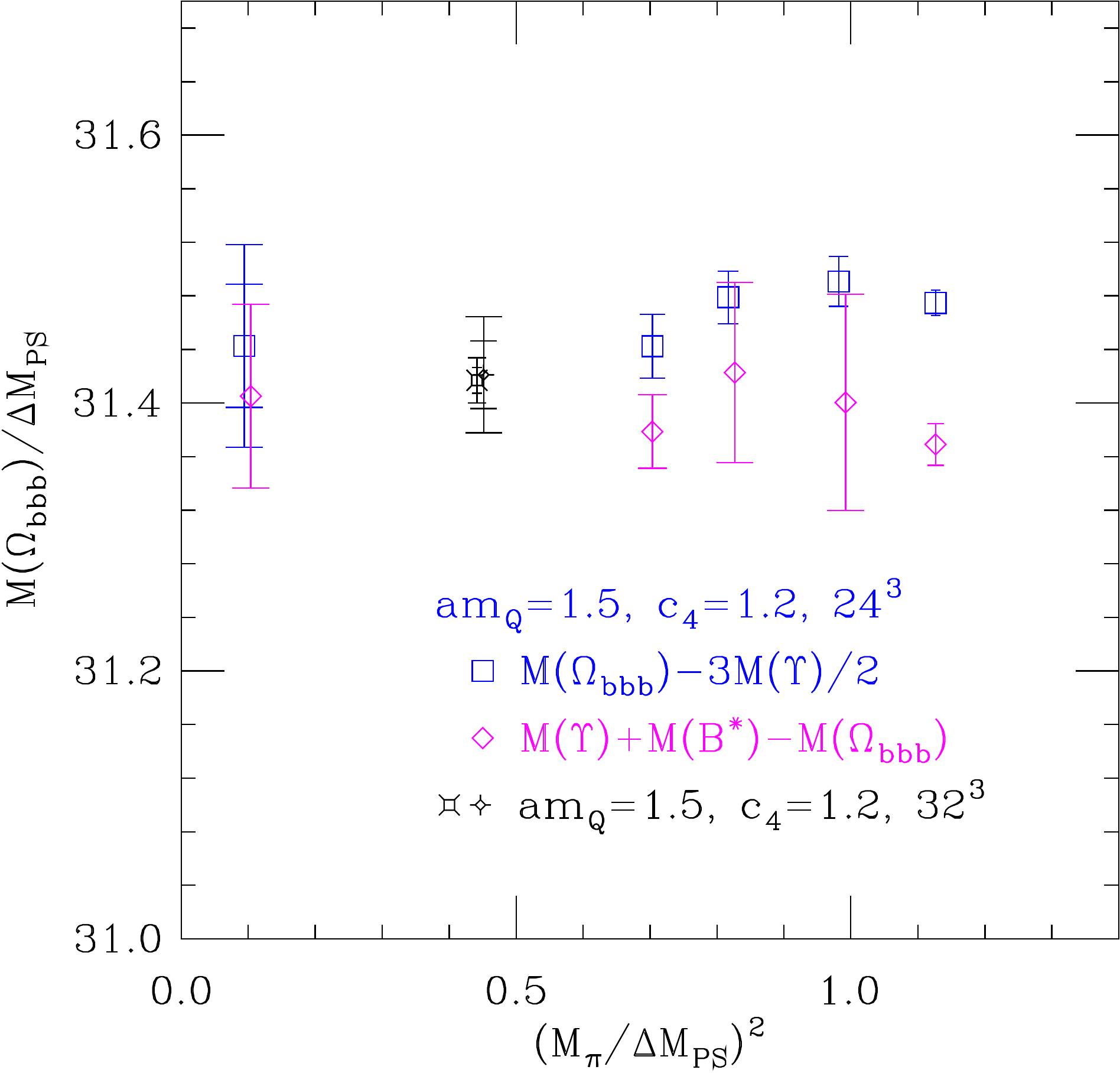}
\end{center}
\caption{
$\Omega_{bbb}$ masses versus $M_\pi^2$.
}
\label{dM_Obbb}
\end{figure}

Figure \ref{dM_Obbb} displays the $\Omega_{bbb}$ mass achieved via 
two schemes: from $E(\Omega_{bbb})-3E(\Upsilon)/2$ and $M(\Upsilon)$; 
or from $E(\Upsilon)+E(B^*)-E(\Omega_{bbb})$ and $M(\Upsilon)+M(B^*)$. 
The two estimates agree in the $N_f=2+1$ chiral limit and on the 
$N_f=2$ lattice.

\begin{table}[!b]
\caption{
Results for $bbc$ and $bcc$ baryon mass splittings (in MeV) 
using $am_Q^{}=1.5$ and $c_4=1.2$ for 
the $N_f=2+1$ chiral limit ($\chi$) and 
the $N_f=2+1$ ensemble {\bf b}. 
The first error comes from the fit, 
the second from the scale setting ($\Delta M_{PS}$).
}
\label{QQc_results}
 \begin{tabular}{lcc}
 \hline
 \hline
 splitting & $N_f=2+1$ ($\chi$) & $N_f=2+1$ ({\bf b}) \\
 \hline
$\Omega_{bcc} - 2B_c + \Upsilon/2$ & 164(27)(12) & 185(9)(6) \\
$\Omega_{bbc} - B_c - \Upsilon/2$ & 177(27)(13) & 179(9)(5) \\
 \hline
$\Omega_{bcc}^* - \Omega_{bcc}$ & 28.8(5.6)(2.2) & 27.2(1.8)(0.8) \\
$\Omega_{bbc}^* - \Omega_{bbc}$ & 21.8(7.0)(1.7) & 26.5(2.2)(0.8) \\
 \hline
$\Omega_{bcc}'^* - \Omega_{bcc}^*$ & 449(41)(34) & 422(15)(13) \\
$\Omega_{bbc}' - \Omega_{bbc}$ & 375(74)(28) & 372(26)(11) \\
 \hline
%$\Omega_{bcc}'^* - \Omega_{bcc}'$ &  &  \\
$\Omega_{bbc}'^* - \Omega_{bbc}'$ & 30(12)(2) & 39.6(4.1)(1.2) \\
% \hline
%$\Omega_{bcc}(2) - \Omega_{bcc}$ &  &  \\
%$\Omega_{bbc}(2) - \Omega_{bbc}$ &  &  \\
 \hline
 \hline
\end{tabular}
\end{table}

For further results (e.g., spin and parity splittings) involving $bcc$ 
and $bbc$ baryons, see Table \ref{QQc_results}. 
Estimates for the first-excited $\Omega_{bbb}$ appear in 
Table \ref{QQQ_results}.

\begin{table}[!b]
\caption{
Results for $bbb$ baryon mass splittings (in MeV) 
using $am_Q^{}=1.5$ and $c_4=1.2$ for 
the $N_f=2+1$ chiral limit ($\chi$),  
the $N_f=2+1$ ensemble {\bf b}, 
and the $N_f=2$ ensemble {\bf a}. 
The first error comes from the fit, 
the second from the scale setting ($\Delta M_{PS}$). 
($^\ddagger$Constant fit, using only the three lightest $M_\pi$ values.)
}
\label{QQQ_results}
 \begin{tabular}{lccc}
 \hline
 \hline
 splitting & $N_f=2+1$ ($\chi$) & $N_f=2+1$ ({\bf b}) & $N_f=2$ ({\bf a}) \\
 \hline
$\Omega_{bbb} - 3\Upsilon/2$ & 179(21)(14) & 172(10)(5) & 167.6(4.4)(3.3) \\
$\Upsilon + B^* ...$ &  &  &  \\
$\;\;- \Omega_{bbb}$ & 162(31)(12) & 150(13)(5) & 169(12)(3) \\
 \hline
%$\Omega_{bbb}(2) ...$ &  &  &  \\
$\Omega_{bbb}(2) ...$ & 211(128)(16) &  &  \\
$\;\;- \Omega_{bbb}$ & 438(33)(33)$^\ddagger$ & 411(45)(12) & 480(64)(9) \\
 \hline
 \hline
\end{tabular}
\end{table}

Using (jackknifed) combinations of four correlators, we were able to 
compare the effect of trading one of the $b$ quarks for a lighter one 
in the mesons with the same effect in the baryons. 
These results appear in Table \ref{QQ-Qq-QQQ+QQq_results}. 
For the first meson difference, $\Upsilon-B_c$, the spin changes from 
1 to 0, but these states are much better determined (experimentally) 
than the corresponding $\eta_b$ and $B_c^*$ (this extra spin jump adds 
about 50 MeV to the first four rows). 
Considering the errors, there appears to be little difference between 
the splittings involving $\Upsilon-B_s^*$ and $\Upsilon-B^*$, but 
exchanging the $b$ quark for a strange or light one appears to have a 
larger effect on the baryons than exchanging $b \to c$, especially when 
looking at baryons with fewer heavy quarks \cite{MakeItShort}.

\begin{table}[!b]
\caption{
Comparison of meson and baryon mass splittings (in MeV) resulting from 
the replacement of one $b$ quark with a lighter one. 
Using $am_Q^{}=1.5$ and $c_4=1.2$ for 
the $N_f=2+1$ chiral limit ($\chi$) and 
the $N_f=2+1$ ensemble {\bf b}. 
The first error comes from the fit, 
the second from the scale setting ($\Delta M_{PS}$).
}
\label{QQ-Qq-QQQ+QQq_results}
 \begin{tabular}{lcc}
 \hline
 \hline
 splitting & $N_f=2+1$ ($\chi$) & $N_f=2+1$ ({\bf b}) \\
 \hline
$(\Upsilon-B_c) - (\Omega_{bbb}-\Omega_{bbc}^*)$ & 45(14)(3) & 42.3(3.5)(1.3) \\
%$(\Upsilon-B_c) - (\Omega_{bbb}-\Omega_{bbc})$ & 13.4(9.2)(1.0) & 16.7(2.6)(0.5) \\
$(\Upsilon-B_c) - (\Omega_{bbc}-\Omega_{bcc})$ & 16.7(9.8)(1.3) & 16.4(3.4)(0.5) \\
%$(\Upsilon-B_c) - (\Omega_{bbc}^*-\Omega_{bcc})$ & $-15$(12)(1) & $-13.5$(4.4)(0.4) \\
$(\Upsilon-B_c) - (\Omega_{bb}-\Omega_{bc}')$ & 17(32)(1) & 18(12)(1) \\
$(\Upsilon-B_c) - (\Xi_{bb}-\Xi_{bc}')$ & 34(36)(3) & 27(13)(1) \\
 \hline
$(\Upsilon-B_s^*) - (\Omega_{bbb}-\Omega_{bb}^*)$ & $-29$(20)(2) & $-29.3$(8.4)(0.9) \\
$(\Upsilon-B_s^*) - (\Omega_{bbc}-\Omega_{bc}')$ & $-91$(27)(7) & $-84.8$(8.8)(2.6) \\
$(\Upsilon-B_s^*) - (\Omega_{bb}-\Omega_{b})$ & $-130$(50)(10) & $-137$(17)(4) \\
$(\Upsilon-B_s^*) - (\Xi_{bb}-\Xi_{b}')$ & $-171$(65)(13) & $-154$(27)(5) \\
 \hline
$(\Upsilon-B^*) - (\Omega_{bbb}-\Xi_{bb}^*)$ & $-35$(20)(3) & $-29.7(7.9)(0.9)$ \\
$(\Upsilon-B^*) - (\Omega_{bbc}-\Xi_{bc}')$ & $-112$(30)(8) & $-96$(10)(3) \\
$(\Upsilon-B^*) - (\Omega_{bb}-\Xi_{b}')$ & $-137$(57)(10) & $-139$(21)(4) \\
$(\Upsilon-B^*) - (\Xi_{bb}-\Sigma_b)$ & $-179$(74)(13) & $-153$(33)(5) \\
 \hline
 \hline
\end{tabular}
\end{table}

\section{Discussion}
\label{SectDiscussion}

% -- spin splittings: $\Upsilon-\eta_b$ slightly low, larger discrepancy for 
% $N_f=2$ ($m_Q$, $c_4$, strange quenched); 
% agreement for known B mesons, but seemingly larger ones for $B^*-B$ than 
% for $B_s^*-B_s$ (different excited-state contamination?); 
% $N_f=2$, $1P-\chi_{b0}$ discrepancy ($m_Q$, $c_4$, strange quenched)
% -- Finite-volume effects: some signs in bottomonia 2S and 1D states, 
% possible strong effects for 2P and 2D; no obvious effects in B mesons
% -- $B_{s0(1)}^*$ below the $B^{(*)}K$ threshold: relatively narrow states, 
% more reliable spin splittings

We reserve this section for a brief discussion of possible sources of 
systematic errors affecting our results \cite{MakeItShort}.

$(1)$ Finite-volume effects: 

As can be seen in Table \ref{latticetable}, all volumes are rather small 
when compared to the dynamical pion masses ($M_\pi L \lesssim 4$), 
with $L$ ranging from 1.7--1.8 fm for the $N_f=2+1$ ensembles to 2.2 fm 
for the $N_f=2$ one. 
Obvious energy enhancements can be seen in some hadrons containing light 
quarks: e.g., $\Lambda_b^{(*)}$, $\Sigma_b^{(*)}$, $\Xi_b^{(',*)}$, and 
possibly even $B^*-B$. 
We can only imagine how large such effects may be for higher excitations. 
However, there are no indications that hadrons containing strange quarks as 
their lightest constituents suffer from these small boxes.

$(2)$ Different previous lattice scales; too heavy $s$ and $c$ quarks: 

During the tuning and running stages of the present ensembles 
\cite{QCDSF_Nf2,QCDSF_Nf2p1}, different scales were used than the 
one in the current study. 
The same was true during a tuning of the charm quark 
\cite{Bali:2011dc,Bali:2012ua}. 
The fact that we now use $\Delta M_{PS}$ from the bottomonia system 
to set the scale leads to a seemingly $\sim 10$\% enhancement of the 
$c$ and $s$ (in the chiral limit) masses.

$(3)$ For $N_f=2$, too heavy $b$ quark: 

Looking at Fig.~\ref{Mkin_1mm}, one can see that on the $N_f=2$ ensemble, 
$am_Q^{}=1.5$ is too large (by $\sim 10$\%) for the bottom quark.

$(4)$ Different smearings; different excited-state ``contamination'': 

On ensembles {\bf c} and {\bf d}, much less smearing was chosen for the 
light and strange quarks in an attempt to better identify excited states. 
However, this may lead to different levels of contamination from excited 
states when considering the ratios of smeared-smeared correlators on the 
different ensembles leading to the $N_f=2+1$ chiral limit.

$(5)$ Relatively heavy $u,d$ quarks; long chiral extrapolations: 

For many systems considered herein, this one is not so much a systematic 
error, outside of the fact that it systematically causes larger errors in 
the chiral limit. 
More trouble arises from this when the system considered should cross a 
threshold between ensemble {\bf b} and the chiral limit.

\section{Conclusions and Outlook}
\label{SectConclusion}

We have presented the heavy-hadron spectrum arising from NRQCD-approximated 
$b$ quarks and improved clover-Wilson $c,s,d,u$ quarks on $N_f=2$ and $2+1$ 
lattices. 
Singly, doubly, and triply heavy ($b,c$) systems were considered and 
results were found for spin splittings, alternate parities, and in some 
cases, ``radial'' excitations. 
Relative mass differences between mesons and baryons resulting from the 
exchange of one $b$ quark for a lighter one were presented as well. 
A number of systematics were pointed out (e.g., small volumes; see above), 
but these have not been fully quantified and the reader must use some 
caution with the results. 
Perhaps more trustworthy are the lower-lying spectra (no radial excitations) 
of hadrons containing only $b$, $c$, or $s$ valence quarks. 
A more careful analysis of all the data generated 
(e.g., with $am_Q^{}=3.0$ or the $40^3$, $N_f=2$ correlators) 
and further runs on existing $32^3$, $N_f=2+1$ ensembles could uncover more 
about the errors incurred, and an expansion of the code \cite{NRcode} could 
lead to a study of possible $b\bar bq \bar q$ states, but we must excuse 
ourselves from taking this project further as we have precious little free 
time and have found new ways in which to be wrong \cite{Burch:2013xna}.

\begin{acknowledgments}
Light-quark propagators were generated with the use of the Chroma \cite{Chroma} 
software library. 
We thank our colleagues for saving the light- and strange-quark propagator 
files on ensembles 
{\bf a}, {\bf b}, and {\bf e} \cite{QCDSF_Nf2,Bali:2011dc} and for making 
them available to the group at large. 
We would like to thank S.~Gutzwiller and R.~Schiel for help in checking that 
we were properly reading the light-quark propagator files and 
J.~Najjar for providing initial plaquette results for our $u_0$ values. 
We would like to thank 
M.~G\"ockeler for helpful suggestions on many topics and 
P.~P\'erez-Rubio for useful discussions involving heavy-hadron interpolators. 
Simulations were performed at the Uni-Regensburg Rechenzentrum and 
Institute for Theoretical Physics and we thank 
the Sch\"afer and Braun Chairs for continued access and 
the administrators for keeping the machines running smoothly. 
This work was supported in part by the DFG (SFB TR-55) 
% gender and family... ha! my ass... 
and mostly by a patient family. 
Beyond the scope of this project, we would like to thank those who still 
deem it worthwhile to discuss physics with the unaffiliated. 
In the end, it must be admitted that the author has been in this place too 
long, rendering him a cynic \cite{Devil}. 
Apologies for where it shows. 
%like here... 
\end{acknowledgments}


\begin{thebibliography}{1234567}
\bibitem{MakeItShort}
  Excuse the curt presentation in places, but this is as much exodus as it 
  is catharsis and the later-written parts of the manuscript will show it. 
  Initially intended as part of a larger study with larger lattices, 
  these results have been sitting around partially analyzed for a while, 
  but they may have retained some relevance. 
  As it always should be, use or damn what you will.
\bibitem{Chroma_qrot}
  Before they can be combined with the nonrelativistic (NR) heavy-quark propagator, 
  the quark propagators produced by the Chroma library \cite{Chroma} must be 
  rotated from an ``alternative'' chiral spin-basis to the NR one: 
  $M_{NR}^{-1} = U M^{-1} U^\dagger$, where 
  $U = \frac{1}{\sqrt{2}} ( {}_{-\sigma_y}^{\;\;\,\sigma_y} \;\; {}_{\sigma_y}^{\sigma_y}  )$.
\bibitem{Oops}
  An error in the flavor projections in our code has left us without the 
  possibility of exploring the $J=\frac12$ states arising from the 
  $\Sigma^2_{Qi}$ operator.
\bibitem{FitResults}
  For the sake of brevity, we leave out such things as effective-mass plots, 
  fit choices, $\chi^2$ values, etc. 
  The interested reader may peruse such details included in the tarball of 
  the code \cite{NRcode}.
\end{thebibliography}
\end{document}